\documentclass[fleqn,usenatbib]{mnras}
\usepackage[T1]{fontenc}
\usepackage{ae,aecompl}
\usepackage{graphicx}
\usepackage{amsmath}
\usepackage{amssymb}
\usepackage{url}
\graphicspath{{figure/}}

\newcommand{\kms}{\,km\,s$^{-1}$}

\newcommand{\Msun}{M$_\odot$}

\newcommand{\pc}[2]{pc$^{-2}$}
\newcommand{\Lya}{Ly$\alpha$}
\newcommand{\fesc}{$f_\text{esc}$}
\newcommand{\fescstar}{$f^{\star}_\text{esc}$}

\newcommand{\nionH}{$\dot{n}_\text{ion,H}$}
\newcommand{\NionHeII}{$\dot{N}_\text{ion,\ion{He}{II}}$}
\newcommand{\nionHeII}{$\dot{n}_\text{ion,\ion{He}{II}}$}
\newcommand{\HII}{\ion{H}{II}}

\newcommand{\HeII}{\ion{He}{II}}
\newcommand{\HeIII}{\ion{He}{III}}
\newcommand{\tauCMB}{$\tau_\text{CMB}$}
\newcommand{\mdot}{$\dot{m}$}
\newcommand{\mBH}{${M}_\text{BH}$}

\defcitealias{Yung2019}{Paper~I}
\defcitealias{Yung2019a}{Paper~II}
\defcitealias{Yung2020}{Paper~III}
\defcitealias{Yung2020a}{Paper~IV}
\defcitealias{Yung2021a}{Paper~VI}

\defcitealias{Somerville2008}{S08}
\defcitealias{Somerville2012}{S12}
\defcitealias{Somerville2015}{SPT15}
\defcitealias{Somerville2015a}{SD15}
\defcitealias{Hirschmann2012a}{H12}
\defcitealias{Popping2014}{PST14}

\defcitealias{Kubota2018}{KD18}
\defcitealias{Shen2020}{S20}

\title[SAM forecasts -- V. AGN and He reionization]{Semi-analytic forecasts for \textit{JWST} -- V. AGN luminosity functions and helium reionization at $z$ = 2--7}

\author[L. Y. A.\ Yung et al.]{L. Y. Aaron\ Yung,$^{1}$\thanks{E-mail: aaron.yung@nasa.gov}\thanks{NPP Fellow}
Rachel S.\ Somerville,$^{2}$ Steven L.\ Finkelstein,$^{3}$ 
\newauthor Michaela\ Hirschmann,$^{4,5}$ Romeel\ Dav\'e,$^{6,7,8}$ Gerg\"{o}\ Popping,$^{9}$ 
\newauthor Jonathan P.\ Gardner$^{1}$ and Aparna Venkatesan$^{10}$
\\
$^{1}$ Astrophysics Science Division, NASA Goddard Space Flight Center, 8800 Greenbelt Rd, Greenbelt, MD 20771, USA\\
$^{2}$ Center for Computational Astrophysics, Flatiron Institute, 162 5th Ave, New York, NY 10010, USA\\
$^{3}$ Department of Astronomy, The University of Texas at Austin, Austin, TX 78712, USA\\
$^{4}$ DARK, Niels Bohr Institute, University of Copenhagen, Lyngbyvej 2, DK-2100 Copenhagen, Denmark\\
$^{5}$ INAF - Astronomical Observatory of Trieste, Via G.B. Tiepolo 11, I-34143 Trieste, Italy\\
$^{6}$ Institute for Astronomy, University of Edinburgh, Edinburgh EH9 3HJ, UK\\
$^{7}$ University of the Western Cape, Bellville, Cape Town 7535, South Africa\\
$^{8}$ South African Astronomical Observatories, Observatory, Cape Town 7925, South Africa\\
$^{9}$ European Southern Observatory, Karl-Schwarzschild-Strasse 2, D-85748 Garching, Germany\\
$^{10}$ Department of Physics and Astronomy, University of San Francisco, 2130 Fulton Street, San Francisco, CA 94117, USA
}

\date{Accepted XXX. Received YYY; in original form ZZZ} 

\pubyear{2021}
\begin{document}
\label{firstpage}
\pagerange{\pageref{firstpage}--\pageref{lastpage}}
\maketitle

\begin{abstract}
Active galactic nuclei (AGN) forming in the early universe are thought to be the primary source of hard ionizing photons contributing to the reionization of intergalactic helium. However, the number density and spectral properties of high-redshift AGN remain largely unconstrained. 
In this work, we make use of physically-informed models calibrated with a wide variety of available observations to provide estimates for the role of AGN throughout the Epoch of Reionization. 
We present AGN luminosity functions in various bands between $z = 2$ to 7 predicted by the well-established Santa Cruz semi-analytic model, which includes modelling of black hole accretion and AGN feedback. 
We then combine the predicted AGN populations with a physical spectral model for self-consistent estimates of ionizing photon production rates, which depend on the mass and accretion rate of the accreting supermassive black hole. 
We then couple the predicted comoving ionizing emissivity with an analytic model to compute the subsequent reionization history of intergalactic helium and hydrogen. 
This work demonstrates the potential of coupling physically motivated analytic or semi-analytic techniques to capture multi-scale physical processes across a vast range of scales (here, from AGN accretion disks to cosmological scales).
Our physical model predicts an intrinsic ionizing photon budget well above many of the estimates in the literature, meaning that helium reionization can comfortably be accomplished even with a relatively low escape fraction. 
We also make predictions for the AGN populations that are expected to be detected in future \emph{James Webb Space Telescope} surveys.
\end{abstract}



\section{Introduction}

Active galactic nuclei (AGN), powered by accreting super-massive black holes (SMBHs), are some of the brightest objects detected in the distant Universe. 
Driven by a set of physical and radiative processes distinct from the ones that power star-forming galaxies, this class of objects is more efficient at producing hard ionizing photons loosely defined as ionizing radiation with energy sufficient to ionize helium (e.g. >54.4eV). 
However, these AGN are also relatively rare compared to star-forming galaxies. 
Furthermore, AGN that are faint or obscured are extremely hard to detect, leaving large uncertainties in their number density and spectral characteristics, which propagate to uncertainties in estimates of the total ionizing photon budget during the Epoch of Reionization (EoR) \citep[e.g.][]{Hopkins2007, Madau2015}.

The bulk of the relatively bright high-redshift AGN were identified in the large, multi-colour Sloan Digital Sky Survey \citep[SDSS;][]{York2000, Richards2002, Jiang2008, Bovy2011} with optical fluxes and colour selection techniques. The more recent Baryon Oscillation Spectroscopic Survey \citep[BOSS;][]{Dawson2013, Ross2013} based on SDSS-III has identified over 20 000 AGN at $2.2 < z < 3.5$. 
And at $z\gtrsim 4$, high-redshift AGN can be selected from \emph{Hubble Space Telescope} (\emph{HST}) deep-field surveys, such as the Cosmic Assembly Near-infrared Deep Extragalactic Legacy Survey \citep[CANDELS;][]{Grogin2011, Koekemoer2011}, with techniques based on rest-frame UV fluxes \citep{Giallongo2015} or follow-up with X-ray capable instruments, such as the \textit{Chandra} Observatory \citep{Weisskopf2000, Nandra2015, Civano2016, Luo2016, Xue2016, Kocevski2017}. 
The ground-based Subaru Telescope \citep[e.g.][]{Matsuoka2018, Matsuoka2019a} and the space-based \textit{X-Ray Multi-Mirror Mission} \citep[\textit{XMM-Newton;}][]{Jansen2001} are also powerful facilities for detecting AGN during EoR.
Combining the rich set of available observations has provided significant insights into the number density and distribution of AGN at high redshift, and how they have evolved across cosmic time \citep{Manti2017, Shen2020}.

The \emph{James Webb Space Telescope} \citep[\emph{JWST};][]{Gardner2006} is the next-generation NASA flagship facility, and will be capable of exploring both AGN and galaxies forming in the primordial universe. 
With its unprecedented infrared (IR) sensitivity, it is expected to detect faint objects several magnitudes below the detection limits of current generation instruments.
And with the photometric and spectroscopic capabilities of its on-board instruments, \emph{JWST} is also well-equipped to identify AGN in high-redshift surveys via conventional colour selection techniques and follow-up spectroscopic diagnostics. 
In addition, the Mid-Infrared Instrument (MIRI) also opens up the possibility of detecting obscured AGN at $z \lesssim 3$ \citep{Yang2021}. 
Studies have also predicted that \emph{JWST} will be able to provide indirect constraints on the BH masses and accretion rates of AGN, and possibly the seeding and growth mechanisms for BHs in primordial halos \citep{Natarajan2017, Volonteri2017, Amarantidis2019}.

Numerical simulations have long been used to investigate and experiment with the physics that drives AGN and their host galaxies. This is especially important for understanding processes that take place on physical scales or time scales that cannot be directly studied with observations. 
Simulations have been conducted in various flavours, including fully-hydrodynamic, cosmological simulations such as \textsc{Simba} \citep{Dave2019,Thomas2021}, Illustris \citep{Vogelsberger2014}, Illustris-TNG \citep{Weinberger2017, Nelson2019}, and Horizon-AGN \citep{Dubois2016, Kaviraj2017, Volonteri2016, Volonteri2020}; `zoom-in' simulations selected from within a larger cosmological volume \citep[e.g.][]{Choi2014, Angles-Alcazar2017a}; and idealized, isolated galaxy simulations \citep[e.g.][]{Hopkins2010}. These simulations incorporate many of the physical processes expected to be important in shaping BH growth, including mechanical and radiative feedback from SMBHs \citep[e.g.][]{Choi2014,Choi2017,Weinberger2017,Su2021}, although many uncertainties remain about the details of how these processes work.
The computational resources required for hydrodynamic simulations increase rapidly with the simulated volume and mass resolution. This tension makes it extremely difficult to simultaneously capture both a large volume and include the full range of relevant physical processes, which take place across a vast range of spatial and temporal scales \citep[e.g.][]{Angles-Alcazar2020}.

On the other hand, semi-analytic and empirical approaches provide flexible, physically motivated alternatives for efficiently modelling large ensembles of objects 
\citep[e.g.][]{Fanidakis2013, Menci2014, Ricarte2018, Dayal2020, Fontanot2020, Orofino2021}. 
Combining the high computational efficiency and the modularized structure of these models, controlled experiments can also be conducted to explore the parameter space and alternative prescriptions for specific physical processes. 
In analytic or semi-empirical studies that focus on the IGM phase transition, even simpler empirical relations between halos and AGN luminosity are adopted \citep{McQuinn2009, Madau2015, Hassan2016, Finkelstein2019, Faucher-Giguere2021}.

The EoR is one of the most significant events in cosmic history, when the intergalactic medium (IGM) transitioned from neutral to ionized \citep{Fan2006, Fan2006a}. 
Although it is unlikely that AGN contributed significantly during hydrogen reionization, AGN are thought to be the main contributor to the reionization of helium, which could have begun as early as $z\sim7$ \citep[e.g.][]{Madau2015, Finkelstein2019}, and does not conclude until $z\lesssim3$, as inferred by the He II Gunn-Peterson effect observed in quasar spectra \citep[e.g.][]{Jakobsen1994, Heap2000, Syphers2013}.

Currently, large uncertainties remain in the ionizing photon budget contributed by AGN at high redshift. This can be broken down into three moving parts: the number density of sources, their spectral characteristics, and the fraction of ionizing photons that were able to escape to the IGM. 
For the relatively bright AGN, the first two have been quite well constrained with the aforementioned photometric and spectroscopic surveys. 
However, both of these remain unconstrained for the fainter populations, which are potentially the main contributors that drove the reionization process at $z\gtrsim4$.

The escape fraction of ionizing photons is the least constrained among the three components and is also notoriously difficult to model. 
Simulations have shown that this quantity is highly sensitive to a large set of intricate, multi-scale geometrical and physical features,
including mass and angular momentum of galaxies \citep[e.g.][]{Paardekooper2011}, density profile and distribution of sources \citep[e.g.][]{Benson2013, Kimm2019, Ma2020}, gas and dust content \citep[e.g.][]{Popping2017a}, and stellar feedback effects \citep[e.g.][]{Kimm2014, Kimm2017, Trebitsch2017}.
Many studies attempting to constrain the escape fraction via observations have arrived at similar conclusions \citep[e.g.][]{Dijkstra2016,Guaita2016,Shapley2016,fletcher2019,Nakajima2020}. 
Although these studies mostly focus on star-forming galaxies, these complications are expected to also apply to AGN, with the added complication that photons must propagate out of the optically thick material that fuels the black hole (torus) as well as through the interstellar medium (ISM) of the host galaxy and circumgalactic medium (CGM). Studies have also shown that the escape fraction can easily have several orders of magnitude of scatter and does not correlate well with any particular global galaxy property \citep[e.g.][]{Ma2015a, Paardekooper2015}.


In this work, we construct a physical, source-driven modelling pipeline that integrates many multi-scale physical processes, ranging from accretion activity near BHs to halo-scale gas accretion and cooling, and explore their collective effects on cosmological-scale processes such as the progression of intergalactic helium reionization. 
The models are calibrated to reproduce low-redshift observations, and are then used to make predictions for AGN populations that are too faint and at too high a redshift to have existing direct observational constraints. 
Our predicted hard ionizing photon emissivity accounts for the combined effects of the cosmological physics-based model for the AGN population and a spectral model that reflects the underlying BH masses and accretion rates. 
By interfacing these model components to an analytic reionization model for intergalactic hydrogen and helium \citep[e.g.][]{Madau1999}, we can make inferences about the ionizing escape fraction required to match the existing constraints. 
The modelling pipeline developed in this work is an extension of the versatile semi-analytic model for galaxy formation, which tracks a wide variety of galaxy formation physics and allows some of our novel predictions to be self-consistently compared to the properties of their host galaxies \citep{Somerville1999, Somerville2008, Hirschmann2012a}.

In this series of \emph{Semi-analytic forecasts for JWST} papers, we present a comprehensive collection of predictions for galaxies and AGN forming in the early universe that are anticipated to be observed by \emph{JWST} and other future facilities. 
In \citealt{Yung2019,Yung2019a} (hereafter \citetalias{Yung2019} and \citetalias{Yung2019a}), we presented detailed predictions for the photometric and physical properties for high-redshift galaxy populations, for which we also provided predictions for how future \emph{JWST} detections can be used to further constrain galaxy formation physics. In \citealt{Yung2020,Yung2020a} (hereafter \citetalias{Yung2020} and \citetalias{Yung2020a}), we modelled the production of ionizing photons by stars in these galaxies and further investigated the subsequent reionization history of intergalactic hydrogen. 
Results from these previous works paint a coherent picture showing that the predicted galaxy population, which is able to reproduce a wide range of observed distributions of $M_\text{UV}$, $M_*$, and SFR up to $z \sim 10$, are also producing sufficient ionizing photons to yield a reionization history that is consistent with all of the existing IGM and CMB constraints.
In this penultimate paper of the series (Paper V), we extend our predictions to include AGN and their contributions to the reionization of intergalactic hydrogen and helium.
All results presented in the paper series will be made available at \url{https://www.simonsfoundation.org/semi-analytic-forecasts-for-jwst/}. Full object catalogues will be released as part of an upcoming, final paper of this series \citepalias[Yung et al., in preparation; or][]{Yung2021a}.

The key components of this work are summarized as follows: the key modelling components are summarized briefly in Section \ref{sec:model}. Predicted AGN characteristics and the resulting helium reionization history are presented in Section \ref{sec:results}. We also investigate the role AGN played during cosmic hydrogen reionization in Section \ref{sec:H_reion}. We discuss our findings in Section \ref{sec:discussion}, and a summary and conclusions follow in Section \ref{sec:snc}.

\section{The Modelling Framework}
\label{sec:model}
In this section, we present the components that make up the fully semi-analytic modelling pipeline for AGN growth and cosmic He reionization. Throughout this work, we adopt cosmological parameters that are consistent with the ones reported by \citeauthor{Planck2016} (XIII \citeyear{Planck2016}): $\Omega_\text{m}=0.308$, $\Omega_\Lambda=0.692$, $H_0=67.8$ km s$^{-1}$Mpc$^{-1}$, $\sigma_8=0.831$, and $n_s=0.9665$. We adopt hydrogen and helium mass fractions $X=0.75$ and $Y=0.25$, respectively.

\subsection{Semi-analytic model for galaxies and AGN}
\label{sec:scsam}
The foundation of this series of papers is the Santa Cruz semi-analytic model (SAM). This modelling framework is described in full in \citet{Somerville1999}, \citet*{Somerville2001}, \citet{Somerville2012}, \citet[][hereafter \citetalias{Popping2014}]{Popping2014}, \citet[][hereafter \citetalias{Somerville2015}]{Somerville2015}, and in particular, the co-evolution model for galaxies, black holes and AGN in \citet[][hereafter \citetalias{Somerville2008}]{Somerville2008}, \citet{Hirschmann2012a}, and \citet{Porter2014}. The free model parameters are calibrated to a subset of $z \sim 0$ observations without recalibration for higher redshifts. See appendix B in \citetalias{Yung2019} and section 2.2.3 in \citet{Somerville2021} for details regarding the calibration criteria and process.

Dark matter halo merger and growth histories, or simply `merger trees', are the backbone of the semi-analytic modelling approach. 
These merger trees can either be extracted from cosmological-scale numerical simulations or constructed using the extended Press-Schechter (EPS) formalism \citep{Press1974, Bond1991, Sheth2002}. Although numerical merger trees are better at capturing the interplay between halos and the cosmic environment in which they are embedded, these simulations are subject to tension between the simulated volume and resolution, and can be computationally expensive. Both a large simulated volume and high mass resolution are required to properly capture the black hole seeding in early, low-mass halos, while sampling rare massive halos with their merger histories sufficiently resolved. 
On the other hand, EPS-based algorithms are able to flexibly generate merger trees for halos of any given masses \textit{on-demand}, which have been shown to reproduce the statistical results for a large ensemble of merger trees extracted from $N$-body simulations \citep{Lacey1993, Somerville1999a, Zhang2008, Jiang2014}.

In this work, we adopt the EPS merger tree algorithm presented in \citet{Somerville1999a} and \citetalias{Somerville2008}. For each output redshift, we create a grid of 200 halo masses that are equally spaced in the range $V_\text{vir} = 100 - 1400$ \kms.
Note that the halo mass range used in this work is adjusted from previous papers in the series to better capture the rare, massive halos that are likely to host SMBH. 
For each of the masses in the grid, we apply the algorithm to construct a hundred Monte Carlo merger history realizations. 
These merger histories trace back to either a minimum progenitor mass of $M_\text{res} = 10^{10}$\Msun\ or 1/100th of the root halo mass, whichever is smaller.
Independent output halo mass grids are created between $z = 2$ to 7 at half redshift increments.
The expected number density of each of these dark matter halos is weighted by the fitted halo mass function in the respective redshift provided by \citet{Rodriguez-Puebla2016} based on results from the MultiDark simulation suite \citep{Klypin2016}.
See appendix C in \citetalias{Yung2019} for details regarding the halo mass function adopted for this work.

The versatile Santa Cruz SAM is well-equipped to capture the symbiotic relationship between AGN and their host galaxies, where the cold gas reservoir that fuels the accreting SMBH is constantly modified by a large collection of cooling and galaxy formation physics, and in turn, the star formation activity in the host galaxy is regulated by AGN feedback. We refer the reader to \citetalias{Yung2019} for a full description of the galaxy formation model configurations adopted in this work and the rest of the series. Here we provide a concise summary of the model components that are related to the seeding and growth of SMBHs only.

\subsubsection{The growth of bulges}

In the Santa Cruz SAM, a basic ansatz is that BH properties are tightly connected to the properties of the \emph{bulge} component of galaxies, as supported by observations. When gas initially cools, it is assumed to accrete into a disc, and stars that form out of that gas are assumed to have disc-like kinematics and morphology. Disc stars can be moved into a bulge component via two mechanisms. The first is mergers: when galaxies merge, if the merger mass ratio is larger than a critical value, all stars from both progenitors are placed into a bulge component. For lower mass ratio mergers, the stars from the lower mass progenitor are deposited in the bulge component of the descendent galaxy. 
This approach is motived by results from numerical simulations of galaxy mergers and is widely used in semi-analytic models. We refer the reader to \citetalias{Somerville2008} and \citet{Hirschmann2012a} for the relevant references and further details.

The second way for bulges to grow is via `disc instabilities'. 
A `disc instability' (DI) mode for bulge formation and growth has been implemented in the Santa Cruz SAM by \citet{Hirschmann2012a, Porter2014}, and a handful of DI model variants have been implemented and tested in these works. 
Here we provide a concise summary of the DI model adopted in this work, which has been shown to be in good agreement with observed galaxy morphologies from $z\sim 2$--0 \citep{Brennan2015}, and 
we refer the reader to the above references for a full description. 
Following early analytic models by \citet{Toomre1964} and \citet*{Efstathiou1982}, a disc becomes unstable when the ratio of dark matter mass to disc mass falls below a critical value
\begin{equation}
    M_\text{disc,crit} \equiv \frac{V_\text{max}^2 r_\text{disc}}{G\epsilon_\text{crit}} \text{,}
\end{equation}
where $V_\text{max}$ is the maximum circular velocity of the halo, $r_\text{disc}$ is the scale length of the cold gas disc, and $M_\text{disc}$ is the combined mass of stars and gas in the disc. 
Therefore, by defining a disc coefficient as follows
\begin{equation}
    \epsilon_\text{disc} \equiv \frac{V_\text{max}}{\sqrt{GM_\text{disc}/r_\text{disc}}} \text{,}
\end{equation}
a disc becomes unstable when $\epsilon_\text{disc} < \epsilon_\text{crit}$. 
Numerical simulations report a range of $\epsilon_\text{crit} \sim 0.6$ to 1.1 depending on the composition of the disc, where discs with higher gas content tend to have a lower instability threshold than the ones with lower gas content.
We adopt $\epsilon_\text{crit} = 0.75$, which is chosen to reproduce the observed bulge fractions as a function of mass in nearby galaxies. 
At each time-step, if the disc is unstable, a fraction of stars and gas is moved from the disc to the bulge to achieve marginal stability, with the cold gas assumed to fuel and be consumed by a starburst. The fraction of gas and stars being moved is proportional to the ratio of cold gas and stars in the disc.

\subsubsection{BH seeding and growth}

A seed black hole of fixed mass $M_\text{seed}$ is assigned to every `top-level' halo in our merger trees.
In our fiducial configuration, we adopt a BH seed mass of $M_\text{seed}=10^4$\Msun. This is somewhat larger than the mass of a seed BH that would be expected to be left behind by massive Population III stars \citep{Abel2002}, but our merger trees do not reach the halo masses that are expected to host these Pop III stars, so this accounts for some previous growth.
Previous testing has shown that most results are not sensitive to BH seed mass values within the range of $M_\text{seed}\sim100$\Msun\ to $10^4$\Msun\ (see \citetalias{Somerville2008}).

Rapid BH growth results in the radiatively efficient `bright mode' of AGN activity. This mode is fuelled by cold gas accretion, usually triggered by mergers and disc instabilities \citep[e.g.][]{Hopkins2007b}. 
In the Santa Cruz SAM, radiatively efficient accretion is triggered when a merger with mass ratio of $\mu > 0.1$ occurs, where $\mu$ is the mass ratio of the baryonic components and the dark matter within the central part of the galaxy (see \citetalias{Somerville2008} for the precise definition). 
The BH in the two progenitor galaxies are assumed to merge rapidly to form a new BH, with mass conserved. 
Motivated by results from simulations presented in \citet{Hopkins2007a}, the post-merger BH will then grow at the Eddington rate until it reaches a critical mass
\begin{equation}
    \log\left(\frac{M_\text{BH,crit}}{M_\text{*,bulge}}\right) = f_\text{BH,crit}[-3.27+0.36\;\text{erf}((f_\text{gas}-0.4)/0.28)]\text{,}
\end{equation}
where $M_\text{*,bulge}$ is the stellar mass of the bulge and $f_\text{gas}$ is the cold gas fraction of the larger progenitor. 
$f_\text{BH,crit}$ is a free parameter that is calibrated such that the output $M_\text{bulge}$--$M_\text{BH}$ relation reproduces the relation observed at $z\sim0$ \citep{McConnell2013}. This critical mass corresponds to the energy needed to halt further accretion and begin to power a pressure-drive outflow. After reaching the critical BH mass, the accretion rate declines as a power law as described in \citep{Hirschmann2012a}.

We also specify an assumed Gaussian scatter in the BH critical mass, which is set to $\sigma_\text{BH}=0.3$ in our fiducial model \citep[e.g.][]{Somerville2021}. 
In addition, we also include an adjusted model with $\sigma_\text{BH}=0.5$, which has been shown in past studies to better reproduce the bright AGN population at high redshift \citep{Hirschmann2012a}. We note that the intrinsic scatter in the $M_\text{bulge}$--$M_\text{BH}$ relation at $z=0$ is approximately 0.3 dex, and it could have been larger at high redshift, as mergers and other processes will reduce the scatter and the scatter is essentially unconstrained at high redshift. Later in this work, we will also show that the additional scatter in the relation is one way to help produce more bright AGN and yield better agreement with the observed number density for the bright AGN populations.

\begin{figure}
    \includegraphics[width=\columnwidth]{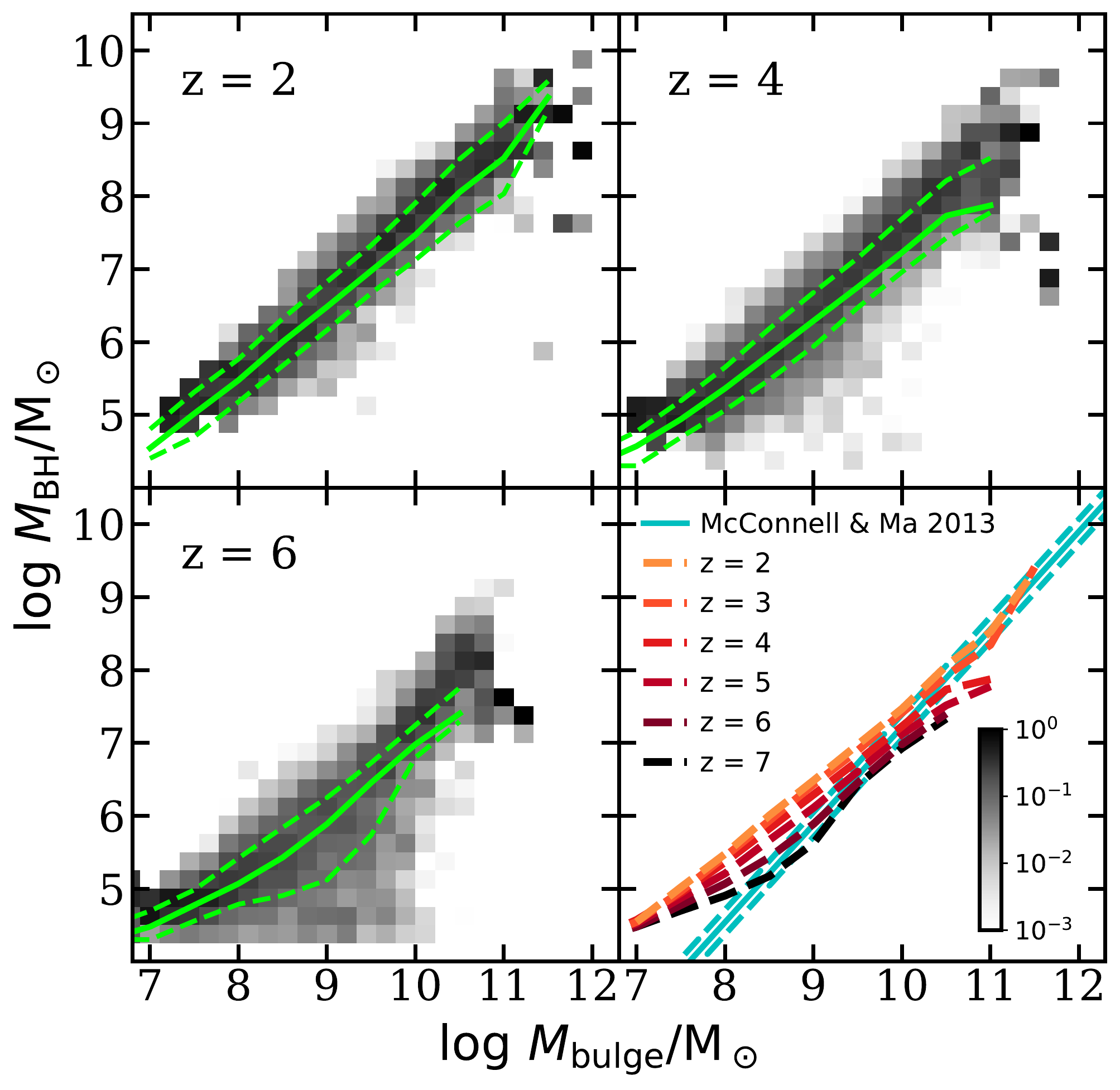}
    \caption{The $M_\text{BH}$--$M_\text{Bulge}$ relation with $\sigma_\text{BH} = 0.3$ predicted by our fiducial model at $z=2$, 4, and 6. The greyscale 2D histograms show the conditional number density per Mpc$^3$, which is normalized to the sum of the number density in its corresponding (vertical) $M_\text{Bulge}$ mass bin.  The green solid and dashed lines mark the 50th, 16th, and 84th percentiles.
        The last panel shows an overlay of the median relation from $z = 2$ to 7. This is compared to the observed constraints from \citet{McConnell2013} at $z=0$, which was used to calibrate our model.}
    \label{fig:mbulge_mBH_z}
\end{figure}

\begin{figure}
    \includegraphics[width=\columnwidth]{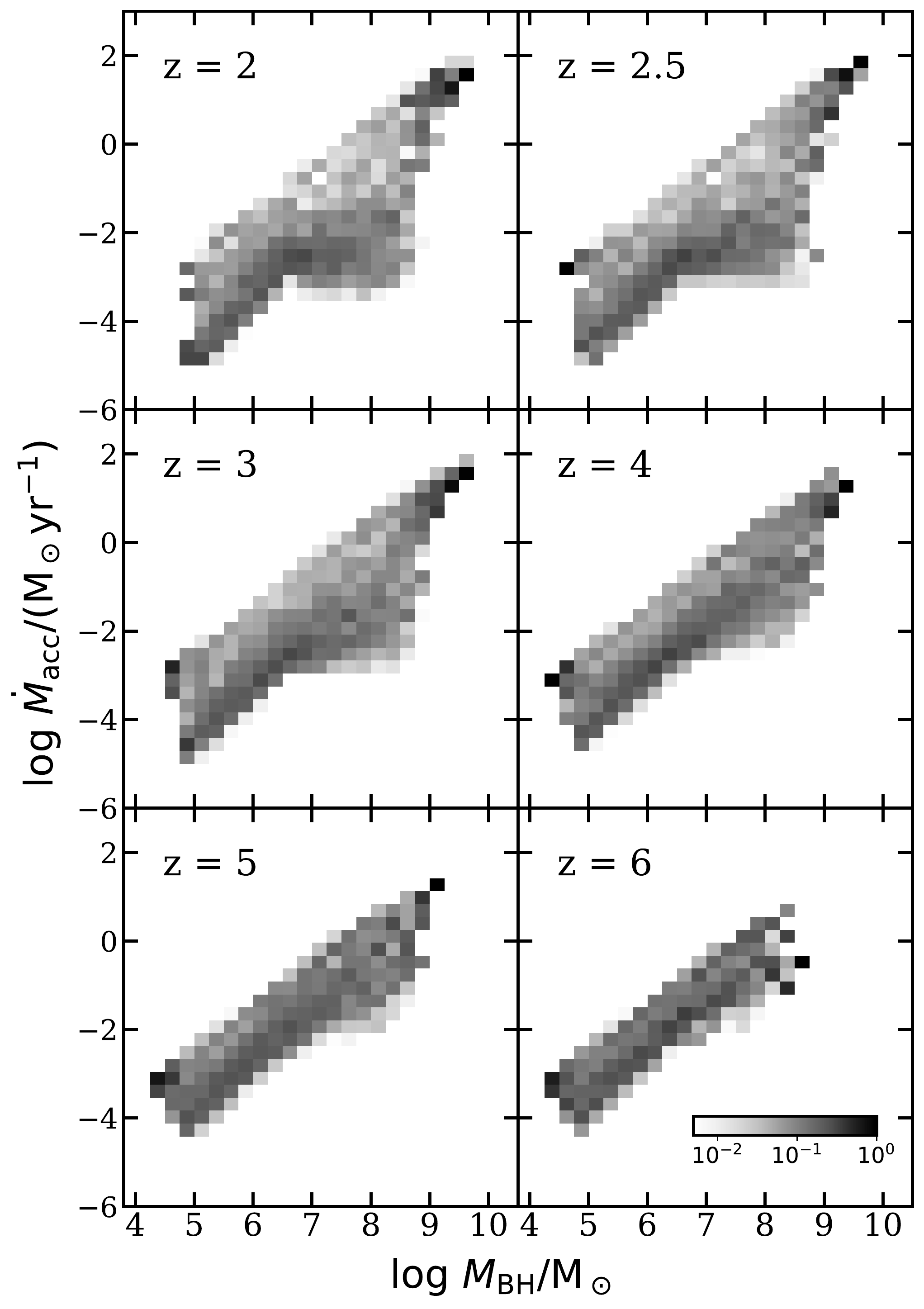}
    \caption{Predicted conditional distribution of black hole mass, \mBH, versus BH accretion rate, $\dot{M}_\text{acc}$, between $z=2$ to 6. The 2D histograms are colour-coded for the conditional number density per Mpc$^3$, which is normalized to the sum of the number density in its corresponding (vertical) \mBH\ bin.}
    \label{fig:BH_diagnostics}
\end{figure}

Rapid BH growth can also follow a disc instability, where we allow a fraction of the mass of the disc $f_\text{BH,disc} = 0.001$ to accrete onto the BH.  This additional fuel is available to be accreted by the BH after a DI event until it is fully consumed.
The DI-associated fueling is not required to respect the critical mass associated with merger-driven fueling, however the timescale for the BH accretion (lightcurve) is given by the same parameters used to define the lightcurve for merger-driven fueling. 
Given both the higher gas fractions \citepalias[e.g.][]{Yung2019a} and gas inflow rates at high redshifts, DI events occur quite frequently and are a major channel for BH growth at early times.
This model is able to reproduce both the observed AGN bolometric luminosity functions and the BH mass--bulge relation observed at $z\sim0$. 
Fig.~\ref{fig:mbulge_mBH_z} shows  our model predictions at $z = 2$ to 7 for the $M_\text{BH}$--$M_\text{bulge}$ relation. 
The additional scatter to low BH masses at $z\gtrsim6$ is likely a result of the lack of time for early forming BH to grow and reach the target relation,  given the short age of the Universe at this redshift.

A secondary, radiatively inefficient `radio' mode driven by Bondi-Hoyle accretion \citep{Bondi1952} in hot quasi-hydrostatic halos with the   isothermal cooling flow model proposed by \citet{Nulsen2000} is also included. 
Both of these accretion modes have been accounted for in the total accretion rate $\dot{M}_\text{acc}$ and the subsequent feedback effects on the ISM and star formation models.
However, the contribution of this mode to the BH accretion rates, especially at high redshift, is negligible. 
As reported in \citetalias{Yung2019}, the feedback effects of this accretion  mode on star formation are also insignificant at $z \gtrsim 4$.

Fig.~\ref{fig:BH_diagnostics} shows the predicted redshift evolution of the joint distribution of $\dot{M}_\text{acc}$ and \mBH\ at $z=2$ to 6. 
The top boundary of the distribution corresponds to the Eddington accretion rate, as our model does not include physical processes that permit accretion at super-Eddington rates. 
The distribution of accretion rates shifts to lower values and the scatter increases towards low redshift with the decline of the merger rate, galaxy gas fractions, and the incidence of disc instabilities.

\subsection{A physical AGN spectral model}
\label{sec:qsosed}

In this work, we adopt the physical AGN spectral model developed by \citet[][hereafter \citetalias{Kubota2018}; also see \citealt*{Done2007} and \citealt{Done2012}]{Kubota2018}.
This model takes BH masses and accretion rates as inputs and produces physically-motivated SEDs spanning a wide energy range.
The \citetalias{Kubota2018} work separately models the emission from three distinct radiative `zones' near an accreting SMBH. These zones include a spherical hot Comptonization region and a standard disc consisting of a warm Comptonization region and a cooler outer disk region. For each of these regions, the emission is modelled assuming blackbody emissivity as described by \citet{Novikov1973}.
In addition, this model accounts for the geometrical self-heating effect due to radiation originating from the hot Comptonization region that subsequently illuminates and reheats the warm Comptonization and outer disc regions. This effect is labelled as `hard X-ray reprocess' in the \citetalias{Kubota2018} model.

The \textsc{qsosed} model form \citetalias{Kubota2018} takes the dimensionless accretion rate defined as \mdot\ $\equiv \dot{M}_\text{acc}/\dot{M}_\text{edd}$, where $\dot{M}_\text{acc}$ is the black hole accretion rate predicted and $\dot{M}_\text{edd}$ is the Eddington accretion rate calculated internally in the SAM based on the mass of the BH. 
For AGN predicted in this modelling pipeline, these quantities are passed to \textsc{qsosed} from the Santa Cruz SAM. 
For the rest of the free parameters, we follow the default configuration from \citetalias{Kubota2018} that is guided by observed spectra of individual nearby AGN. These free parameters correspond to electron temperature $(kT_\text{e}$), spectral index ($\Gamma$), and outer radius ($R$) for the hot and warm Componization components, and a handful of others parameters that characterize the scale height and radii of the disc component (see table 3 in \citetalias{Kubota2018}). 
The calibration of these free parameters is guided by observed high-resolution AGN spectroscopy measurements \citep{Mehdipour2011, Mehdipour2015, Jin2012, Jin2012a}. 
Note that these spectra are either corrected for galactic interstellar reddening or do not suffer from strong reddening effect in the first place. For this reason, dust attenuation and obscuration effects will be modelled separately when these SEDs are being forward-modelled to observable quantities (e.g. UV1450 luminosities).

\begin{figure}
    \includegraphics[width=\columnwidth]{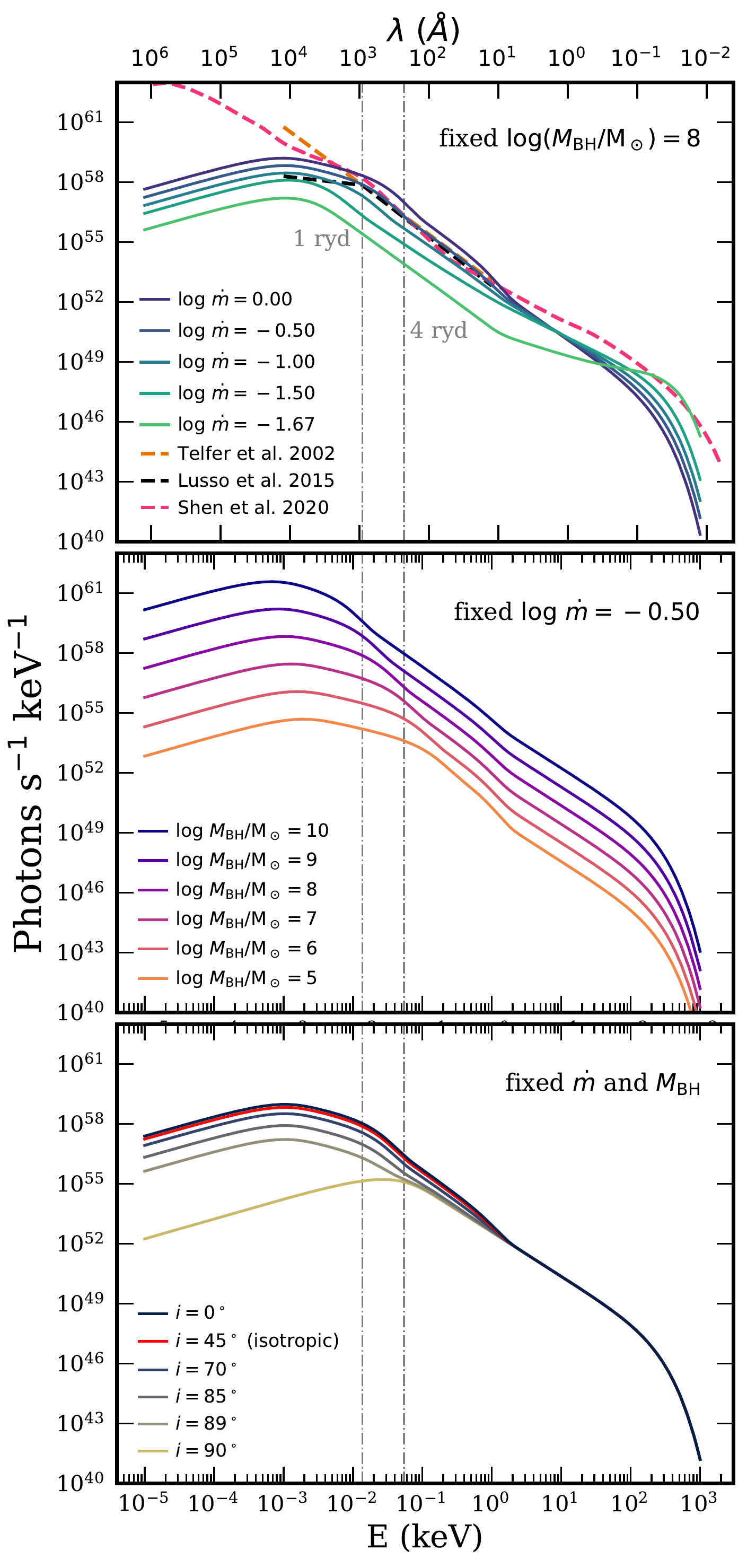}
    \caption{Rest-frame AGN SEDs predicted with \textsc{qsosed} for a range of input Eddington-normalized black hole accretion rates, \mdot, black hole masses, \mBH, and inclinations, $i$. In addition, we also show the wavelength in angstrom corresponding to the energy (top axis). The \textit{top panel} shows predictions from a range of $\log \dot{m} = [0,-1.67]$ assuming a fixed $m_\text{BH} = 10^8$ \Msun; and the \textit{middle panel} shows SEDs for a range of $\log \dot{M}_\text{BH} / \text{M}_\odot = [5, 10]$ assuming a fixed \mdot\  $=-0.5$. The \textit{bottom panel} shows a range of inclination $i = [0^\circ, 90^\circ]$ for a fixed \mdot\ and \mBH. We also mark the energy levels at 1 and 4 ryd to guide the eye. We also show examples of a broken power law with $\alpha_\nu = -1.7$ and $-0.61$ for $\lambda \leq 912$ and $\lambda > 912$, respectively (\citealt{Lusso2015}, black dashed line); a simple power law for $\alpha_\nu = - 1.57$ (\citealt{Telfer2002}, orange dashed line); and the mean SED template adopted in \citet{Shen2020}. See text for details.}
    \label{fig:QSOSED_diagnostics} 
\end{figure}

In Fig.~\ref{fig:QSOSED_diagnostics}, we show a range of sample output AGN spectra from \textsc{qsosed} for a range of fixed values of \mdot\ and \mBH\ computed for an energy grid spanning $10^{-5}$ to $10^3$ keV. 
We show a case where we explore the sensitivity of the SED model to BH accretion rate by showing outputs for a range of $0>\log\dot{m}>-1.67$ for a fixed $\log(m_\text{BH}/\text{M}_\odot) = 8$. Similarly, we also show a similar experiment with BH masses by showing outputs for a range of $5 < \log M_\text{BH}/\text{M}_\odot < 10$ for a fixed $\dot{m} = -0.5$. 
Furthermore, \textsc{qsosed} also takes inclination of the accretion disc as a free parameter, on which UV emission from the disc has a strong dependence, as shown in the bottom panel of Fig.~\ref{fig:QSOSED_diagnostics}. 
For the rest of this work, we adopt a fixed inclination of 45$^{\circ}$ for an overall averaged, isotropic SED for emission quantities such as the emissivity of ionizing photons. On the other hand, for observable quantities, such as band luminosities, we randomly assign an inclination of $0^{\circ} < i < 90^{\circ}$ to account for the possible distribution of inclinations along our line-of-sight.
We note that the inclination of the disc has very little effect in the energy range that corresponds to that of hard ionizing radiation, which originates from the Hot and Warm Comptonization regions.

\begin{figure}
    \includegraphics[width=\columnwidth]{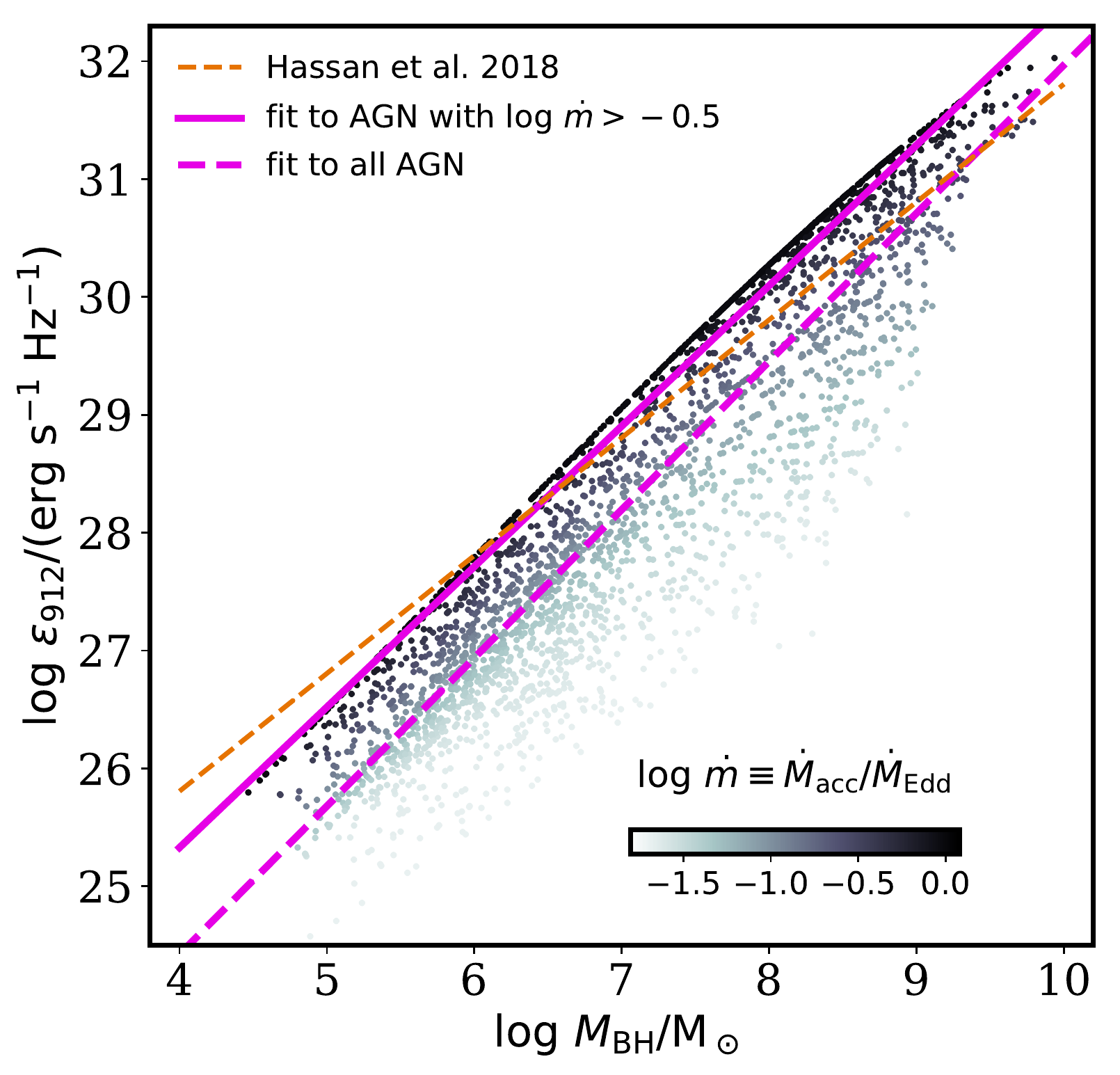}
    \caption{Distribution of \mBH\ versus specific UV emissivity at 1 ryd for our simulated AGN populations, colour-coded by the Eddington-normalized accretion rate. Here we show AGN from an output at $z=3$; however, this relation does not explicitly depend on redshift. The solid magenta lines show the scaling relation fitted to a subset of AGN with high accretion rates $\log\dot{m}>-0.5$, and the dashed magenta line is the fit to all AGN. The orange dashed line shows the empirical relation adopted in \citet{Hassan2018}. See text for details. }
    \label{fig:eps912_compare} 
\end{figure}

For comparison, we add two cases of power-law spectra, $L_\nu = L_{912} (\nu/\nu_{912})^{\alpha_\nu}$, adopted by similar studies. We include a single power law spectrum with $\alpha_\nu = -1.57$ as reported by \citet{Telfer2002}, as adopted in \citet{Hassan2018} and \citet{Haardt2012}, and a broken power law reported by \citet{Lusso2015} with $\alpha_\nu = -1.7$ for at $\lambda \leq 912$\AA\ and $\alpha_\nu = -0.61$, as adopted in the \citet{Shen2020} and \citet{Madau2015} ionizing emissivity calculation. Both of these assume a black hole mass of $m_\text{BH} = 10^8$\Msun and are converted to $L_{912}$ based on the prescriptions used in \citet{Hassan2018}. We also include a composite SED adopted by \citet{Shen2020}, which is normalized to $\nu L_\nu \approx 45.5 \text{ erg s}^{-1} $ at 2500\AA. We note that the deviation  from the \citeauthor{Shen2020} model in the low energy range is due to the lack of dust emission in the \citetalias{Kubota2018} spectral model.

In Fig.~\ref{fig:eps912_compare}, we illustrate the scatter in the specific emissivity at 1 ryd, $\epsilon_{912}$, for the AGN populations predicted by the Santa Cruz SAM when coupled with \textsc{qsosed}. Here we show the $z = 3$ simulated population as an example, but these quantities do not explicitly depend on redshift.
We add a case where $\epsilon_{912}$ is calculated based on black hole masses using a set of empirical conversions, similar to the procedure used in \citet{Hassan2018}, which adopts relations between \mBH\ to $B$-band luminosity $L_B$ and between $L_B$ to $L_{912}$ \citep{Schirber2003, Choudhury2005}.
It is important to note that while the calculation with simple power-laws is in broad agreement with predictions from a physical spectral model, such an approach doesn't capture the scatter introduced by black hole accretion rate, which is shown not to be correlated with black hole mass. 
Here we provide a fit for this relation that takes the form of $\log \epsilon_{912} = A \log M_\text{BH} + B$ fitted with a non-linear least square method. 
For the high accretion rate population with $\log \dot{m} > -0.5$, we find best-fit parameters $A = 1.19$ and $ B = 20.56$, which is fairly similar to the empirical relation derived from observations of luminous AGN. 
However, we find that the inclusion of lower accretion rate AGN populations shifts the relation down to $A = 1.26$ and $B = 19.39$. 
Thus empirical relations derived from observations may misrepresent the full population of AGN due to possible selection biases.

We also note that with the default dissipated energy from the hot Comptonization region set at $L_\text{diss,hot} = 0.02\;L_\text{edd}$, the spectral index for the hot Comptonization region, $\Gamma_\text{hot}$, decreases rapidly for $\log\dot{m} \lesssim -1.7$, which becomes incompatible with the disc-corona geometry assumed by the \citetalias{Kubota2018} model.
Given that observational evidence has suggested that the physical properties of these accreting AGN can change drastically across some accretion rate threshold, leading to a drastic drop in radiative efficiency \citep[e.g.][]{Trump2011}, for the rest of this work, AGN with $\log\dot{m} \lesssim -1.7$ are assumed to be radiatively inefficient.
While the SAM does predict that a population of such low accretion rate BHs exists, we exclude these objects from spectral modelling and their contribution to luminosity functions and ionizing photon budgets. We refer the reader to \citet{Hirschmann2014} for a more in-depth discussion of radiatively inefficient BHs.

The \textsc{qsosed} model is accessed through the spectral fitting package \textsc{xspec}\footnote{\url{https://heasarc.gsfc.nasa.gov/xanadu/xspec/}, v12.11.1} \citep{Arnaud1996}. Compton emission at each annulus is carried out with the \textsc{nthcomp} algorithm \citep{Zdziarski1996, Zycki1999} provided as part of \textsc{xspec}. The original code in \texttt{fortran} is ported to work in a \texttt{python} environment with the \texttt{f2py} tool \citep{Peterson2009}. Other miscellaneous calculations are carried out using functions from  \texttt{astropy} \citep{Robitaille2013, Price-Whelan2018}, \texttt{numpy} \citep{VanderWalt2011}, and \texttt{scipy} \citep{Virtanen2020}.

\subsection{Analytic model for cosmic helium reionization}
\label{sec:reion}

In this section, we present the set of equations that comprise the analytic model that tracks the cosmic helium reionization history.
As discussed in \citet{Wyithe2003} and \citet{Haardt2012}, even though the ionization front of singly ionized helium may briefly overtake that of hydrogen at very early times (e.g. $z\sim10$), the expansion of the hydrogen ionization front is not inhibited by the presence of helium simply because most of the ionizing photons produced are only able to ionize hydrogen (< 24.6 eV). 
Therefore, the following restrictions, $Q_\text{\HeIII} \leq Q_\text{\HII}$ and $Q_\text{\HeII} \leq Q_\text{\HII} - Q_\text{\HeIII}$, are valid and are often assumed in analytic models. 
This allows us to model the reionization of intergalactic helium independently of the hydrogen reionization.

\begin{figure}
    \includegraphics[width=\columnwidth]{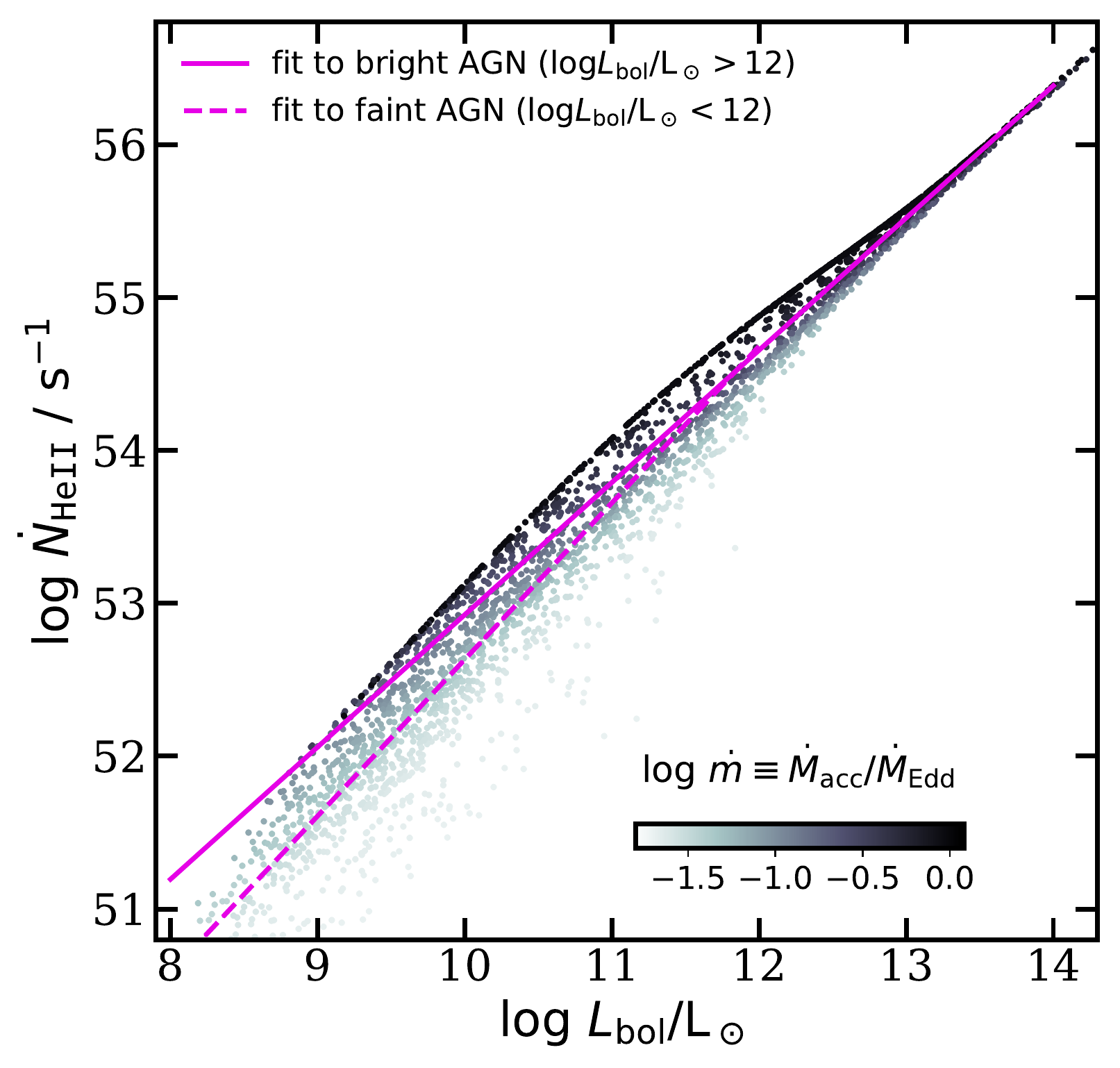}
    \caption{Distribution of $L_\text{bol}$ versus \NionHeII\ for our simulated AGN population. Here we show AGN from at output at $z=3$; however, this relation does not explicitly depend on redshift, colour-coded by the Eddington-normalized accretion rate. 
    The solid magenta line shows the scaling relation fitted only to the bright AGN population with $\log L_\text{bol}/\text{L}_\odot > 12$, and is plotted as an extrapolation below this limit. The dashed magenta line is the fit to AGN with $\log L_\text{bol}/\text{L}_\odot < 12$. This illustrates how adopting relations based on the brightest AGN may over-predict the mean ionizing photon production rate for the full AGN population.}
    \label{fig:Lbol_NHe_scaling} 
\end{figure}

Taking an approach similar to that described in \citetalias{Yung2020}, we calculate the production rate of helium ionizing photons from AGN by integrating the physically modelled spectra. This approach has been shown to be better at capturing subtle changes in the physical properties of the underlying sources, in this case BH mass and accretion rate distributions of the AGN population. 
This approach also provides a realistic estimate of scatter in the AGN properties that is not captured with empirical or power-law conversions.
The production rate of \ion{He}{II} ionizing photons, \NionHeII, from an individual AGN is calculated by integrating its SED at energies above 54.4 eV ($\lambda$ < 228 \AA)
\begin{equation}
    \dot{N}_\text{ion,\HeII} = \int^{\infty}_{\nu_\text{228}} L_{\nu,\text{AGN}} (h\nu)^{-1}d\nu \text{.}
    \label{eqn:NionHeII}
\end{equation} 
Here, $L_{\nu,\text{AGN}}$ is the AGN spectrum from the \textsc{qsosed} model presented (see Section \ref{sec:qsosed}), based on \mBH\ and \mdot\ predicted by the Santa Cruz SAM (see Section \ref{sec:scsam}). Similar to Fig.~\ref{fig:eps912_compare}, in Fig.~\ref{fig:Lbol_NHe_scaling} we show the scaling relation of $L_\text{bol}$ versus \NionHeII\ for the predicted AGN populations at $z=3$, as the model components involved in this calculation do not have any explicit redshift dependence. 
We show that bolometric luminosity is a better tracer for \NionHeII, where the relation is relatively tight. 
Scatter in this relation arises mainly from differences in the spectrum in the UV, which is dominated by the dependence on \mdot\ (see top panel in Fig.~\ref{fig:QSOSED_diagnostics}). 
Fitting these data points with a non-linear least square method with a functional form $\log \text{\NionHeII} = A \log(L_\text{bol}) + B$, we find the best-fit parameters for $A = 0.87$ and $B = 44.27$ for the bright AGN populations with $\log L_\text{bol}/\text{L}_\odot > 12$, which extends into the faint AGN and reproduces well the population accreting near the Eddington rate. 
We also show the case for $\log L_\text{bol}/\text{L}_\odot < 12$, where the relation is slightly shifted, best-fitted with $A = 1.02$ and $B = 42.39$. This steeper relation is caused by the additional scatter in the faint population from AGN with a range of BH accretion rate.

Combining the predicted AGN number density and their spectral properties, we calculate the comoving \ion{He}{II} ionizing emissivity, \nionHeII, by summing the contribution across the predicted AGN populations at each output redshifts 
\begin{equation}
    \dot{n}_\text{ion,\HeII} = \sum_i n_{\text{h},i}\; \dot{N}_{\text{ion,\HeII},i}\; f_{\text{esc,He},i} \text{.}
    \label{eqn:nionHeII}
\end{equation}
Here, $n_\text{h}$ is the number density per Mpc$^3$ for each AGN $i$, assigned based on the virial mass of the host halo. 
Last, $f_\text{esc,He}$ is the escape fraction of helium ionizing photons specifically for AGN. For the remainder of this work, we refer to the AGN-specific escape fraction as \fesc\ unless otherwise specified. 
For the rest of this work, we refer to this quantity as the ionizing photon production rate when assuming \fesc = 1.00, which is to be distinguished from the emissivity that accounts for the amount of ionizing radiation trapped or dissipated in the ISM and CGM, as well as the the potential dust and neutral gas obscuration occurring in the vicinity of the AGN (e.g. nuclear- or torus-scale obscuration). We also note that the physical processes governing the escape fraction of radiation from AGN can be quite different from the ones affecting the escape fraction of radiation from stellar populations.

The comoving \ion{He}{II} ionizing emissivity is then forward modelled into the \textit{volume-averaged} ionizing volume-filling fraction of doubly ionized helium, $Q_\text{\HeIII}$, with the temporal evolution described by the following first-order differential equation
\begin{equation}
    \frac{dQ_\text{\HeIII}}{dt} = \frac{\dot{n}_\text{ion,\HeII}}{\bar{n}_\text{He}} - \frac{Q_\text{\HeIII}}{\bar{t}_\text{rec,\HeIII}}
    \label{eqn:He_reion} \text{,}
\end{equation}
as derived in \citet{Madau1999}. The two terms separately account for the growth of ionized volume and the sink of ionized intergalactic \ion{He}{III} due to recombination.
The growth term is the ratio of the comoving \ion{He}{II} ionizing emissivity, \nionHeII, and the volume averaged comoving number density of intergalactic helium, $\bar{n}_\text{He}$. Here, we adopt IGM mean hydrogen density $\bar{n}_\text{H} = 1.9\times10^{-7}$ cm$^{-3}$ \citep{Madau2014} and the conversion $\bar{n}_\text{He} = \bar{n}_\text{H}\; Y/4(1-Y) = 1.58\times10^{-8}$ cm$^{-3}$.
The sink term is characterized by the recombination of \ion{He}{III} with free electrons by taking the ratio of $Q_\text{\HeIII}$ and the recombination time-scale of \HeIII. 
The recombination time-scale of intergalactic helium is given by
\begin{equation}
    \bar{t}_\text{rec,\HeIII} = [C_\text{\HeIII} \; \alpha_\text{B,\HeII}(T/Z^2) \; (\bar{n}_\text{H}+2\bar{n}_\text{He}) \; (1+z)^3\;Z]^{-1} \text{,}
    \label{eqn:He_recombin}
\end{equation}
where $\alpha_\text{B,\HeIII}(T)$ is the case B recombination coefficient for \ion{He}{III} \citep{Hui1997} and $C_\text{\HeIII}$ is the \HeIII\ clumping factor. 
In this work, we follow \citet{Finkelstein2019} in assuming an IGM temperature of $T = 2\times10^4$ K and $C_\text{\HeIII} = C_\text{\HII}$. For $C_\text{\HII}$, we adopt the redshift-dependent clumping factor from the radiation-hydrodynamical simulation L25N512 by \citet{Pawlik2015}, in which $C_\text{\HII}$  evolves from $\sim1.5$ to $\sim4.8$ between $z\sim14$ to $z\sim6$. At $z \lesssim 6$, the clumping factor is extrapolated using a polynomial fit such that the clumping factor continue to increase approximately linearly to a value of $\sim 6.8$ at $z\sim2$.
The helium reionization history, $Q_\text{\HeIII}(z),$ is then obtained by solving equation (\ref{eqn:He_reion}) using \texttt{scipy.integrate.odeint} and \texttt{astropy.cosmology} \citep{Robitaille2013, Price-Whelan2018}.

\section{Results}
\label{sec:results}

In this section, we present the predictions of our fully semi-analytic, source-driven modelling pipeline. Our results are organized in two parts. (1) the predicted bolometric and band-specific luminosity functions compared to observational constraints and simulated results from numerical hydrodynamic simulations. (2) The ionizing photon production rate for the predicted AGN populations and the subsequent implications for the cosmic helium reionization history.

\subsection{AGN luminosity functions}
One-point distribution functions provide an effective overview for the statistical characteristics of specific observable or physical property among a large ensemble of objects. The distribution functions for luminosity in specific bands are commonly referred to as luminosity functions (LFs).

\subsubsection{Hard X-ray luminosity function}
\begin{figure}
    \includegraphics[width=\columnwidth]{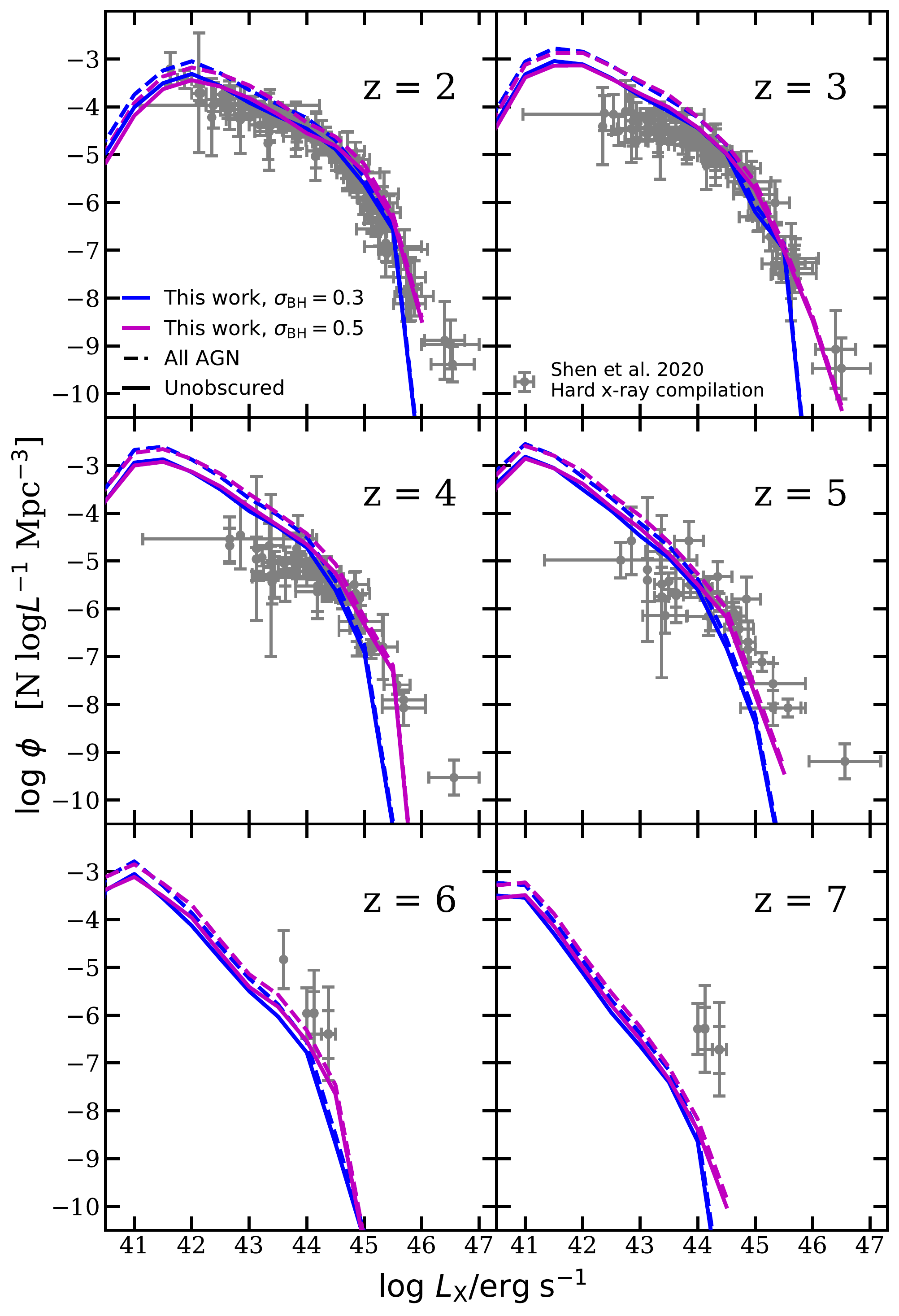}
    \caption{Predicted AGN HXLFs and their evolution with redshift. The lines show the results from our fiducial model ($\sigma_\text{BH} = 0.3$; blue) and adjusted model ($\sigma_\text{BH} = 0.3$; purple). 
    The dashed line shows the full predicted AGN population, while the solid line shows only the unobscured AGN population.
    The data points show the collection of hard X-ray observations compiled by \citetalias{Shen2020}.}
    \label{fig:Compare_xray_LFs}
\end{figure}

In Fig.~\ref{fig:Compare_xray_LFs}, we show the hard X-ray luminosity functions (HXLFs) at $z = 2$--7, where the rest-frame hard X-ray luminosity, $L_\text{X}$, is obtained by integrating the AGN spectra between 2--10 keV.
We consider the hard X-ray LF as the most robust calibration for our models, as we can fully forward model our predictions to this plane, and this quantity is relatively insensitive to obscuration.  
We show predictions made with our fiducial configuration with $\sigma_\text{BH} = 0.30$ and the adjusted configuration with $\sigma_\text{BH} = 0.50$, and for each we show both cases for all predicted AGN and for Compton-thin, \textit{unobscured} AGN, using the $L_X$-dependent obscuration fraction reported by \citet{Ueda2014}.
We show results for all predicted AGN and for only Compton-thin AGN, where we adopted the Compton-thick fraction reported by \citet{Ueda2014}. 
In this exercise, we also gain insight into how the AGN population across a wide luminosity range is affected by distinct physical processes. 
For instance, the number density of the bright AGN is more sensitive to the scatter in the $M_\text{BH}$--$M_\text{bulge}$ relation, as shown  in \citet{Somerville2009}, because of Eddington bias. 
Conversely, the fainter populations are more affected by obscuration, as the fraction of X-ray absorbed AGN generally decreases towards higher intrinsic hard X-ray luminosity \citep{Ueda2003, Ueda2014, Buchner2015}. 
In addition, the faint end slope is affected by the decay timescale and slope of the AGN lightcurves.

Our predictions are compared to a large compilation of hard X-ray observations \citep[e.g.][]{Ueda2003, Ueda2014, Aird2015, Aird2015a, Miyaji2015, Khorunzhev2018} compiled by \citet[][hereafter \citetalias{Shen2020}]{Shen2020}. 
It is very encouraging that the combined AGN sources and spectral models are able to reproduce the observed evolution of high-redshift AGN in the hard X-ray up to $z \lesssim 6$. 
However, it is known that it is challenging to produce the observed number density of very luminous AGN at very high redshift ($z \gtrsim 6$) without assuming either more massive seeds or super-Eddington accretion. It is therefore not surprising that our models also show this discrepancy.

\begin{figure}
    \includegraphics[width=\columnwidth]{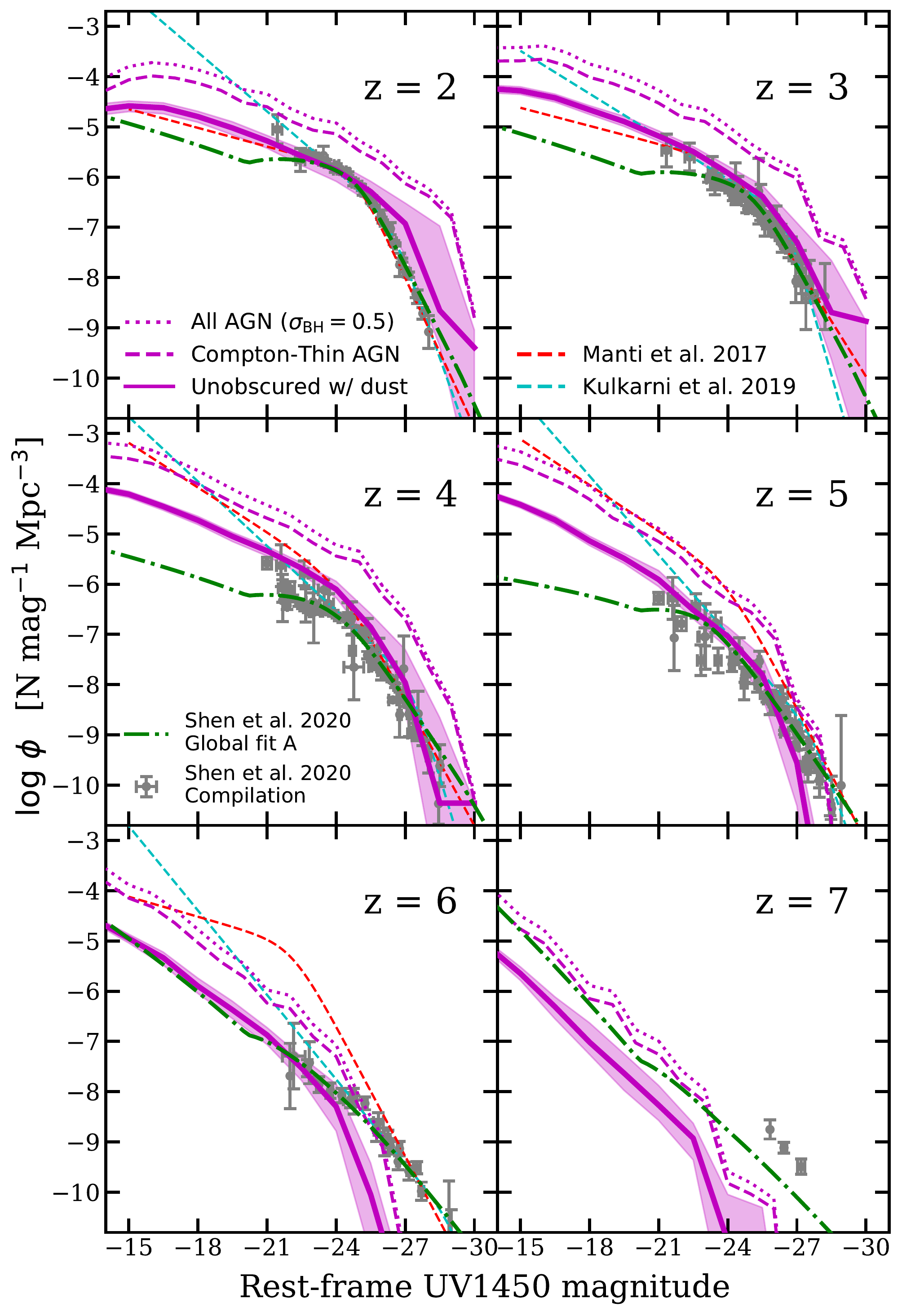}
    \caption{Predicted AGN UVLFs and their evolution with redshift. 
    The lines show the results from our fiducial model ($\sigma_\text{BH} = 0.3$; blue) and adjusted model ($\sigma_\text{BH} = 0.3$; purple). 
    The dashed line shows the full predicted AGN population, while the solid line shows only the unobscured AGN populations.
    The data points show a collection of rest-frame UV1450 observations compiled by \citetalias{Shen2020}. The green dot-dashed line shows the Global fit A of \citetalias{Shen2020}. We also show fits to observed AGN UVLFs reported by \citet[][red dashed line]{Manti2017} and \citet[][cyan dashed line]{Kulkarni2019}. }
    \label{fig:Compare_UV1450_dust_LFs}
\end{figure}

\subsubsection{UV luminosity function}
\label{sec:UVLF}
Similarly, Fig.~\ref{fig:Compare_UV1450_dust_LFs} shows the predicted AGN rest-frame UV luminosity functions (UVLFs) between $z = 2$ to 7, where the luminosity is computed by integrating the SED with a tophat filter of width of 200\AA\ centred at 1450\AA\ with inclination dependence included. See bottom panel in Fig.~\ref{fig:QSOSED_diagnostics} and associated text for a description. 
Similar to the HXLFs, we show both our fiducial and adjusted models, and  we show all AGN and only the Compton-thin ones. For each modelled AGN, an obscured fraction is estimated based on its hard X-ray luminosity and is deducted from the predicted number density per Mpc$^3$.

To forward model our predictions for a direct comparison with observations, we estimate the effect of nuclear scale UV attenuation using the radiation-lifted torus model proposed by \citet{Buchner2017}. This model provides an estimate (drawn stochastically from a distribution) of the hydrogen column density $N_\text{H}$, based on the AGN X-ray luminosity $L_\text{X}$ and \mBH, and statistically reproduces the observed fraction of Compton-thick and Compton-thin AGN. We then assume a mean ratio of total neutral hydrogen to colour excess $N_\text{H}/E(B-V) = 5.8 \times 10^{21} \text{cm}^{-2} \text{mag}^{-1}$ and a total $V$-band attenuation $R_V \equiv A_V/E(B-V) \sim 5$ for AGN to convert the predicted $N_\text{H}$ to $V$-band attenuation $A_V$ \citep{Bohlin1978, Draine2003, Gaskell2007}. This is then converted to attenuation at 1450\AA\ assuming an AGN attenuation curve from \citet{Gaskell2007}. The effect of obscuration and the associated uncertainties are further discussed in Appendix~\ref{appendix:a}. Our predicted attenuation correction leads to a considerable decrease in the predicted number density of observable AGN, of an order of magnitude or more. Because the sampled distribution of $N_\text{H}$  is quite broad (see \citet{Buchner2017} and references therein), this leads to a large variability in the attenuation correction and hence on the predicted number density of AGN, especially at the bright end. 
As there are only a small number of very luminous AGN (both in our modeled sample and in the observed Universe), drawing from this population will not fully sample the distribution of $N_\text{H}$. In order to obtain more robust estimates of the attenuation corrected UVLF, we repeat the random draws of $N_\text{H}$ over 500 independent realizations of our modeled AGN sample, and report the median and the 16th and 84th percentiles in Fig.~\ref{fig:Compare_UV1450_dust_LFs}.

These predictions are compared to a large compilation of observational estimates of the AGN UV LF  \citep[e.g.][]{McGreer2013, McGreer2018, Ross2013, Akiyama2018, Yang2018b, Wang2019}, compiled by \citetalias{Shen2020}. 
In addition, we show fits to  observational results from \citet{Manti2017} and \citet{Kulkarni2019}, as well as the `global fit A' model from \citetalias{Shen2020} for comparison. We note that the faint end of the UVLFs remains largely unconstrained, especially at high redshift. We regard the agreement of our predictions with the available observations as encouraging, although we note that the uncertainties in the obscuration corrections are very large, and could easily result in an order of magnitude uncertainty in AGN number density. However, here we have only included the attenuation correction from the torus scale, neglecting the contribution from the host galaxy scale, which is also expected to be significant \citep[][and references therein]{Buchner2017}. We also show a comparison of these predicted AGN UVLFs to star-forming galaxies UVLFs between $z = 4$ to 7 in Fig.~\ref{fig:UV_LFs_extended}.

\begin{figure*}
    \includegraphics[width=2.1\columnwidth]{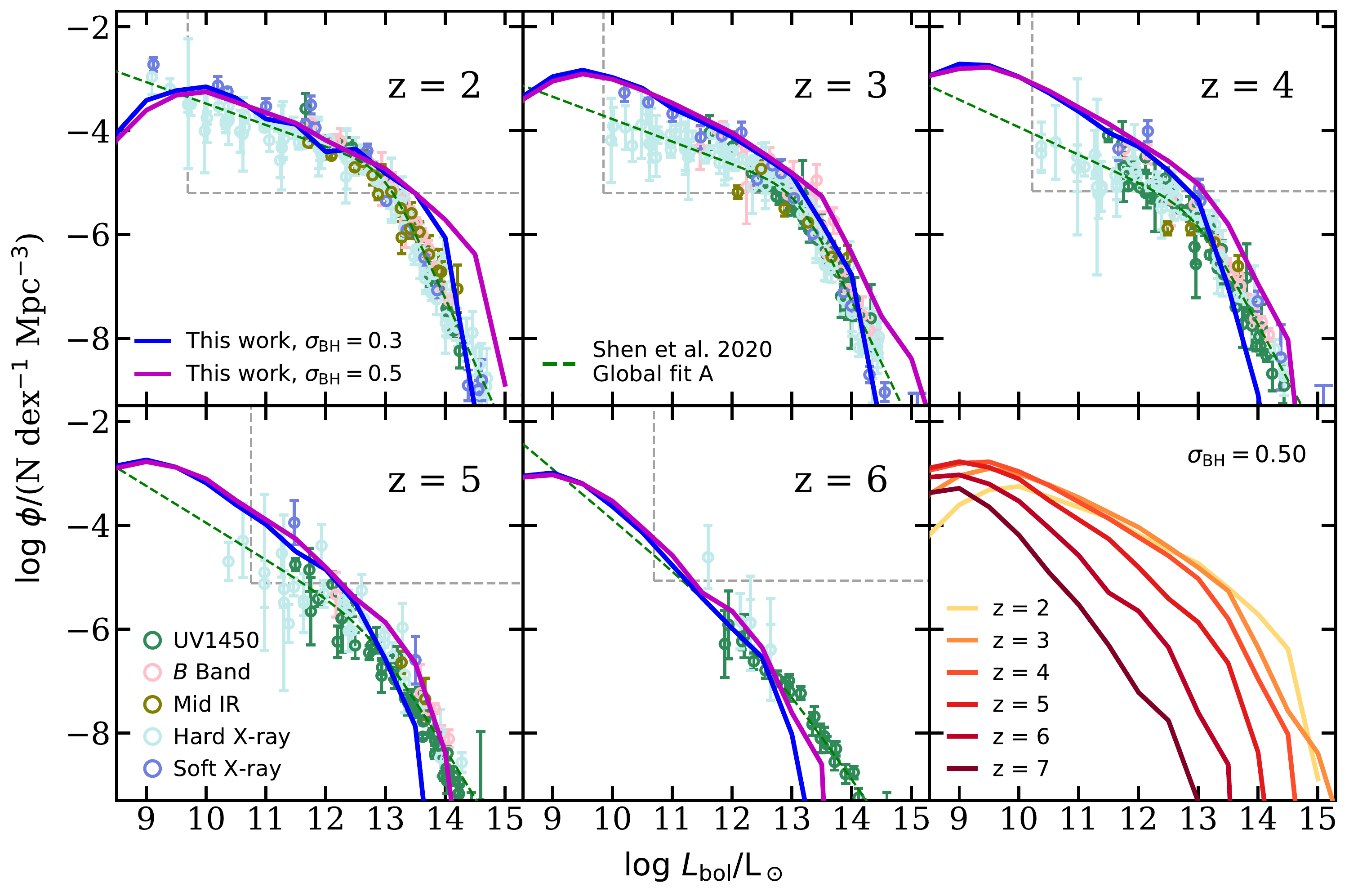}
    \caption{Predicted bolometric AGN luminosity functions and their evolution with redshift between $z=2$ to 7 for our fiducial model ($\sigma_\text{BH}=0.3$; blue line) and adjusted model ($\sigma_\text{BH}=0.5$; purple line). 
        These predictions are compared to multi-wavelength observational constraints compiled by \citet{Shen2020}, which include bolometric corrections and dust and extinction corrections. 
        We also show the fits to the bolometric LFs from \citetalias{Shen2020} (global fit A; green dashed line). 
        The grey dashed lines in each panel indicate the approximate range of objects that are expected to be detected in a typical \emph{JWST} wide-field NIRCam survey, where the vertical boundary loosely corresponds to the detection limit of $m_\text{F200W,lim} = 28.60$ and the horizontal line approximately represents where we expect to detect one object for a $\sim100$ arcmin$^2$ survey area in a redshift slice of $\Delta z = 0.5$.
        The last panel summarizes the predicted evolution of the bolometric LFs in the adjusted model.
        See text for full descriptions of individual elements included in this plot. }
    \label{fig:AGN_Bol_LFs}
\end{figure*}

\begin{figure}
    \includegraphics[width=\columnwidth]{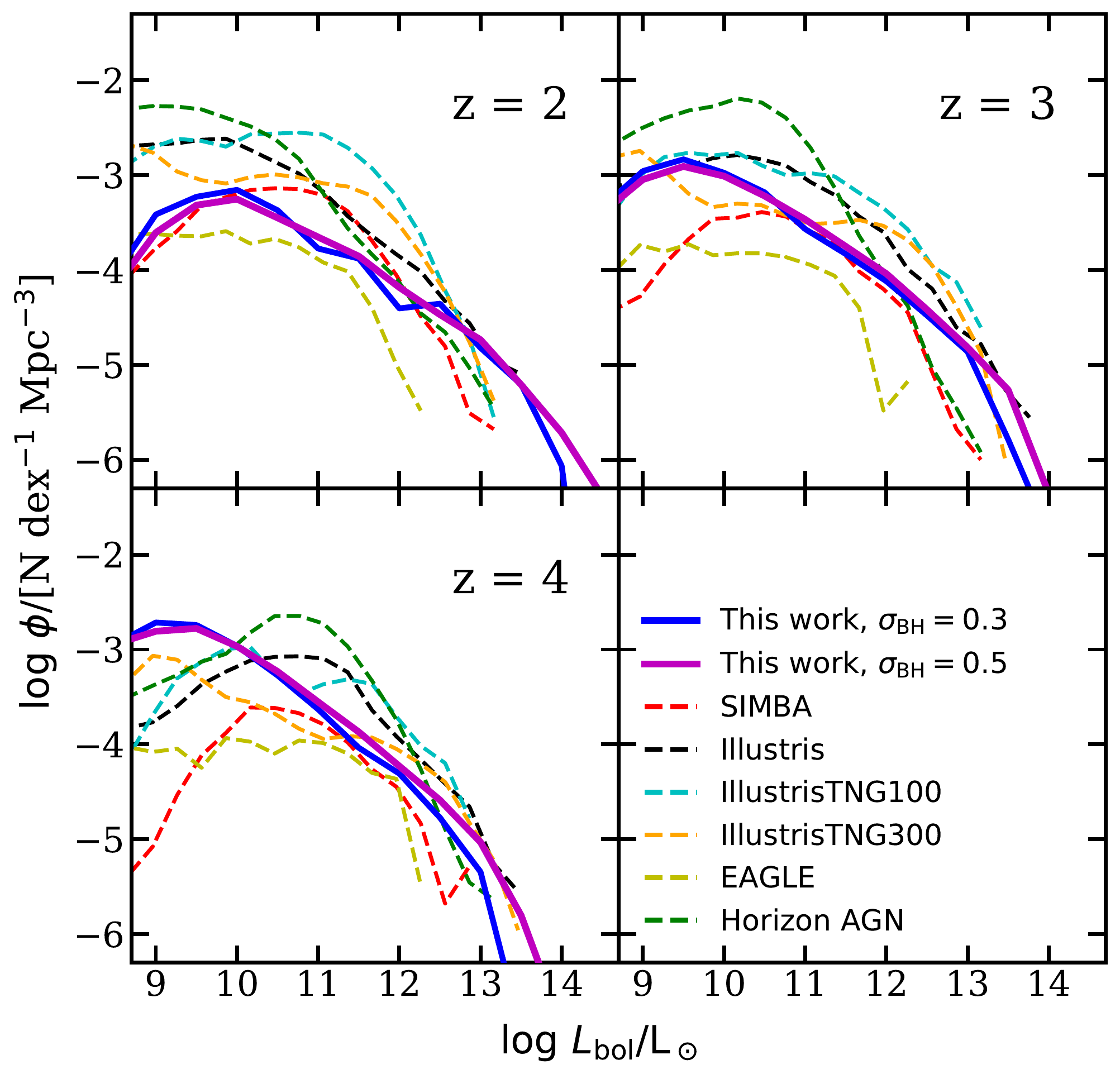}
    \caption{The predicted AGN bolometric luminosity functions from our fiducial model ($\sigma_\text{BH} = 0.3$; blue) and adjusted model ($\sigma_\text{BH} = 0.5$; purple) at $z=2$, 3, and 4, compared to a compilation of modern cosmological hydrodynamic simulations, including \textsc{Simba} \citep{Dave2019}, TNG--100 anf TNG--300 \citep{Nelson2019}, EAGLE \citep{Schaye2015}, Horizon AGN \citep{Dubois2014}, and Illustris \citep{Genel2014}. See text for details. }
    \label{fig:AGN_Bol_LFs_models}
\end{figure}

\subsubsection{Bolometric luminosity function}

The AGN bolometric luminosity function is a useful space in which to compare different theoretical predictions, and is a useful way to consolidate constraints from many different wavelengths. However, it requires more assumptions to get from the observed quantities to the bolometric luminosity function.  For all predicted AGN, we compute the bolometric luminosity, $L_\text{bol}$, by integrating over the full spectrum for each individual object. In Fig.~\ref{fig:AGN_Bol_LFs}, we show the predicted AGN bolometric LFs at $z = 2$ to 7, made with our fiducial configuration ($\sigma_\text{BH} = 0.30$) and the adjusted configuration ($\sigma_\text{BH} = 0.50$). 
These predictions are compared to a large compilation of constraints derived from observations across a variety of wavelengths. In addition to the aforementioned hard X-ray and UV constraints, this comparison also includes optical \citep[e.g.][]{Shen2012, Palanque-Delabrouille2013, Palanque-Delabrouille2016}, mid-IR \citep[e.g.][]{Assef2011,Singal2016}, and soft X-ray \citep[e.g.][]{Hasinger2005, Ebrero2009}. 
\citetalias{Shen2020} converted these multi-wavelength observations to bolometric luminosity by constructing a mean SED template based on observed IR and X-ray SEDs \citep{Hopkins2007,Krawczyk2013} and UV spectral slopes \citep{Lusso2015}.
Note that the UV/optical segment of the constructed SED from \citet{Krawczyk2013} is not strongly affected by dust reddening nor obscuration.
A correction for obscuration effects from gas and dust is applied \citep{Pei1992, Morrison1983, Ueda2014}.
We show the best-fitting AGN bolometric LF reported by \citetalias{Shen2020} that is fitted to all available constraints and does not restrict the evolutionary pattern for the faint-end slope (labelled as `global fit A').

Keeping in mind the significant uncertainties involved in the conversions to bolometric luminosity (which are not reflected in the error bars), we find that our model predictions are very consistent with the available observational constraints. It is encouraging that our fiducial model, which is only calibrated to a subset of observations at $z\sim0$ and was shown to reproduce a wide variety of observed constraints on galaxy properties from the CANDELS survey \citep{Somerville2021}, is also able to reproduce the observed AGN LFs across a wide luminosity range and their evolution across a wide range of redshift. 
It also shows that the scatter in the $M_\text{BH}$--$M_\text{bulge}$ relation has a significant effect on the bright end of the AGN LF, as will any other process that introduces scatter into the AGN luminosity. As noted before, the failure of the models to reproduce the brightest AGN at very high redshift is unsurprising, as the formation of these very rare, massive objects may involve physical processes that are not currently included in this model. See Section \ref{sec:discussion} for a full discussion. 
The binned luminosity functions presented in this work are available in Table~\ref{table:AGN_LF} in Appendix \ref{appendix:d}.

Similar to the illustrations presented in \citetalias{Yung2019} and \citetalias{Yung2019a}, we mark the volume limit and magnitude limit for which objects are expected to be detected in upcoming \emph{JWST} wide-field surveys. 
The volume limits, marked by the horizontal lines, correspond to the a number density where one object is expected to be found in a typical wide-field survey with area of $\sim100$ arcmin$^2$, with a redshift slice of $\Delta z = 0.5$ centred at the output redshift. 
This provides a very rough estimate for when objects become too rare to be expected to be found by \textit{JWST}.
However,  AGN tend to form in more clustered regions, which is not fully represented in our \textit{volume-averaged} approach to estimating object number density.
On the other hand, the vertical lines indicate an approximate detection limit of $m_\text{F200W,lim} = 28.6$, calculated assuming an exposure time comparable to past \textit{HST} wide-field surveys. 
This observed-frame IR magnitude limit is converted to rest-frame UV at each output redshift for the predicted galaxy populations based on the results presented in \citetalias{Yung2019}. 
This rest-UV limit is then translated to $L_\text{bol}$ based on a subset of predicted AGN selected within a narrow bin around the limiting $M_\text{UV}$. 
We note that identifying high-redshift AGN requires selection methods very different from the ones for galaxies \citep{Volonteri2017}. 
This exercise simply provides a rough estimate for which AGN populations are above the detection limit in \textit{JWST} galaxy surveys.

In Fig.~\ref{fig:AGN_Bol_LFs_models}, we compare the same bolometric LFs to the results from a selection of state-of-the-art cosmological hydrodynamic simulations, including \textsc{Simba} \citep{Dave2019}, Illustris \citep{Genel2014}, IllustrisTNG--100 and IllustrisTNG--300 \citep{Nelson2019}, the Evolution and Assembly of GaLaxies and their Environments (EAGLE) simulations \citep{Schaye2015}, and Horizon--AGN \citep{Dubois2014}. The results were compiled by Habouzit et al. (in preparation) and we refer the reader to that work and \citet{Habouzit2021} for in-depth discussion of the simulations and a comparison with observations.
These simulations have comparable simulated volume and mass resolution, and are optimised to simulate the formation of galaxies and AGN in a cosmological volume.
For instance, \textsc{Simba} and Horizon AGN are 100 $h^{-1}$Mpc on a side, with particle mass of the order of $\sim 10^{8}$ \Msun. Illustris, IllustrisTNG--100, and EAGLE are $\sim100$ Mpc on a side, with particle mass of $\sim 10^7$ \Msun.
IllustrisTNG--300 has the largest simulated volume among all compared simulations of 302.6 Mpc on a side, with a relatively coarse dark matter particle mass of $5.9\times 10^7$\Msun.
These simulations each adopt very different prescriptions and parametrizations for baryonic physics, and have adopted different calibration strategies, such as which observational constraints and redshifts to prioritize for the calibration, as well as what redshift range to cover. 
For example, some simulations may terminate at high redshifts and are not calibrated to observational constraints from the local universe.
The bolometric luminosities for AGN in these numerical simulations are calculated based on the predicted black hole accretion rate assuming the standard relation $L_\text{bol} = \epsilon_{r} \dot{M}_\text{acc} c^2 $, where $\epsilon_{r}$ is the radiative efficiency. 
The radiative efficiency is set to $\epsilon_{r} = 0.25$ for Illustris and TNG--100; and $\epsilon_{r} = 0.11$ for Horizon-AGN, EAGLE, and \textsc{Simba}. These parameters are chosen to be consistent with the modelling choices made in the simulations. See \citet{Habouzit2021} and Habouzit et al. (in preparation) for a detailed discussion.
We note that some studies adopt an alternative parametrization that replaces $\epsilon_{r}$ with $\epsilon_{r}/(1-\epsilon_{r})$. 
We also compute $\epsilon_{r}$ based on predictions from our physically-grounded modelling pipeline and compare to values assumed in previous studies in Appendix \ref{appendix:c}.
From this comparison, we find that our predictions are in broad agreement with numerical simulations, with some simulations having steeper faint end slopes and higher AGN number densities, and others having shallower faint end slopes and lower AGN number densities than our models predict.  We note that the numerical simulations probably underestimate the number density of low-luminosity AGN, particularly at high redshift, due to their limited mass resolution, and may have large uncertainties in the number density of luminous AGN, due to their limited volume. Overall,  it is encouraging that our SAM-based predictions are consistent with those from numerical hydrodynamic simulations.

\begin{figure}
    \includegraphics[width=\columnwidth]{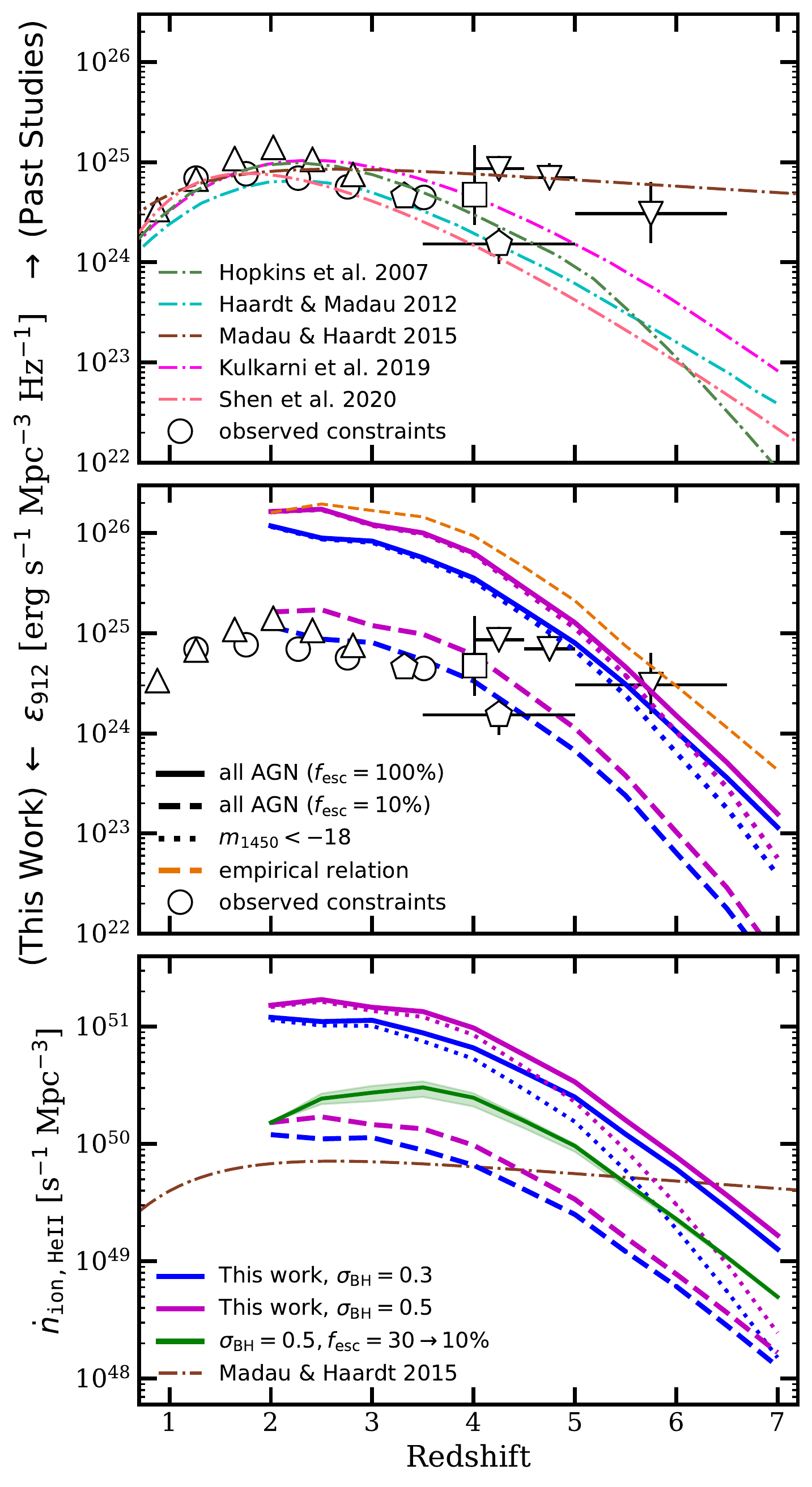}
    \caption{\textit{Top and middle panels:} Redshift evolution of the cosmic specific emissivity at 1 ryd, $\epsilon_{912}$, between $z=2$ to 7 predicted by our fiducial model ($\sigma_\text{BH} = 0.3$; blue) and adjusted model ($\sigma_\text{BH} = 0.5$; purple). The solid line shows all AGN and the dashed line only includes objects with $M_\text{UV} \leq -18$, similar to the other studies shown in this comparison.
    The orange dashed line labelled `empirical relations' shows the emissivity calculated using the \citet{Hassan2018} scaling relation as shown in Fig.~\ref{fig:eps912_compare}. 
    Our results are compared to past analytic studies by \citet{Madau2015} and estimates from fits to observed AGN UVLFs \citep{Hopkins2007, Haardt2012, Shen2020}. We also show a compilation of observations, including \citet[][circles]{Bongiorno2007}, \citet[][squares]{Glikman2011}, \citet[][pentagons]{Masters2012}, \citet[][triangles]{Palanque-Delabrouille2013}, \citet[][upside-down triangles]{Giallongo2015}. \textit{Bottom panel}: \ion{He}{II} ionizing emissivity, \nionHeII, calculated by integrating individual AGN SEDs at $>4$ ryd assuming an escape fraction of unity. We show additional cases with constant \fesc=0.10 (purple and blue dashed lines), and with \fesc\ evolving as a function of redshift 0.30 at $z\gtrsim7$ to 0.10 at $z\sim2$ (green line). The shaded regions correspond to a range of evolutionary paths for the escape fraction. See text for details. }
    \label{fig:AGN_emissivity}
\end{figure}

\subsection{Hard ionizing photon emissivity and helium reionization history}

The reionization of intergalactic helium is a cosmological-scale phase transition driven mainly by AGN. 
Currently, huge uncertainties remain in the estimates of the helium ionizing photon budget, which determines the subsequent progression of the cosmic helium reionization history. 
In this subsection, we present estimates for the total helium ionizing photon budget available during the EoR, based on the predicted AGN number density and spectral characteristics from our physics based models.

In Fig.~\ref{fig:AGN_emissivity}, we show the comoving specific 1 ryd (non-ionizing) photon budget, $\epsilon_{912}$, and the comoving \HeII\ ionizing photon budget ($\lambda < 228$\AA), \nionHeII, as a function of redshift. Here \nionHeII\ is calculated with equation (\ref{eqn:NionHeII}) assuming an escape fraction of unity and no obscuration, to account for all ionizing photons produced by AGN.
These estimates are essentially a combination of the results from the AGN populations from Fig.~\ref{fig:AGN_Bol_LFs} and the individually predicted emissivity shown in Fig.~\ref{fig:eps912_compare} and Fig.~\ref{fig:Lbol_NHe_scaling}. 
We show results from both the fiducial ($\sigma_\text{BH} = 0.3$) and adjusted  ($\sigma_\text{BH} = 0.5$ model, and find that the additional population of luminous AGN in the adjusted model can contribute nearly a factor of two more ionizing photons towards lower redshifts. The tabulated ionizing photon budget is provided in Table \ref{table:nionHeII}. 
We also note that $\epsilon_{912}$ may not be a very good tracer for \nionHeII, given that high-redshift AGN tend to accrete at a higher accretion rate than their low-redshift counterparts with comparable black hole masses, which may have an impact on the overall spectral slope and introduce an effective redshift-evolution in the ratio between UV and bolometric luminosities.

\begin{figure}
    \includegraphics[width=\columnwidth]{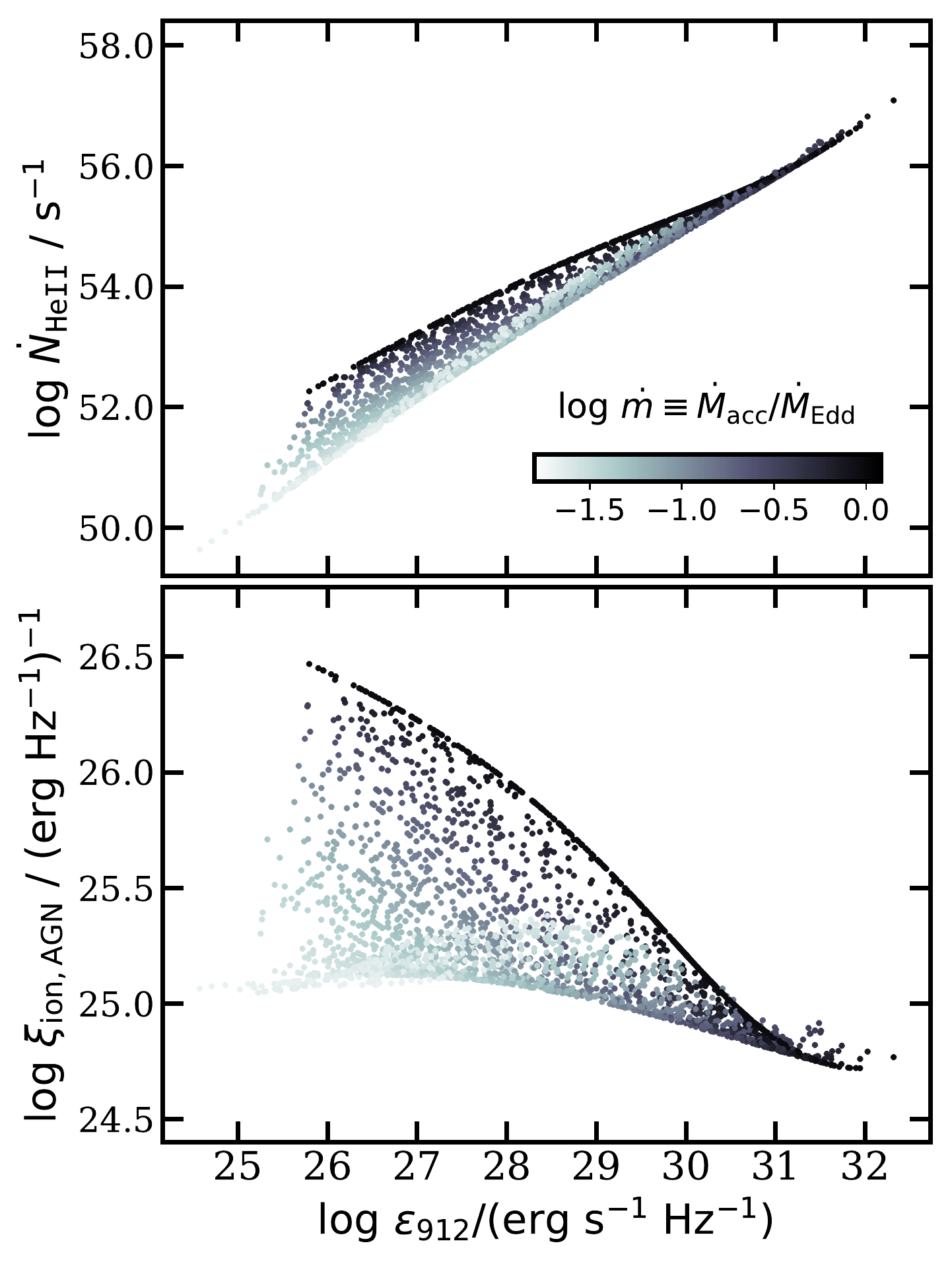}
    \caption{Distribution of $\epsilon_{912}$ versus \NionHeII\ (\textit{top panel}) and $\xi_\text{ion,AGN}$ (\textit{bottom panel}) for our simulated AGN population, colour-coded by the Eddington-normalized accretion rate. Here we show AGN from at output at $z=3$; however, this relation does not explicitly depend on redshift. }
    \label{fig:eps912_nHe}
\end{figure}

\begin{figure}
    \includegraphics[width=\columnwidth]{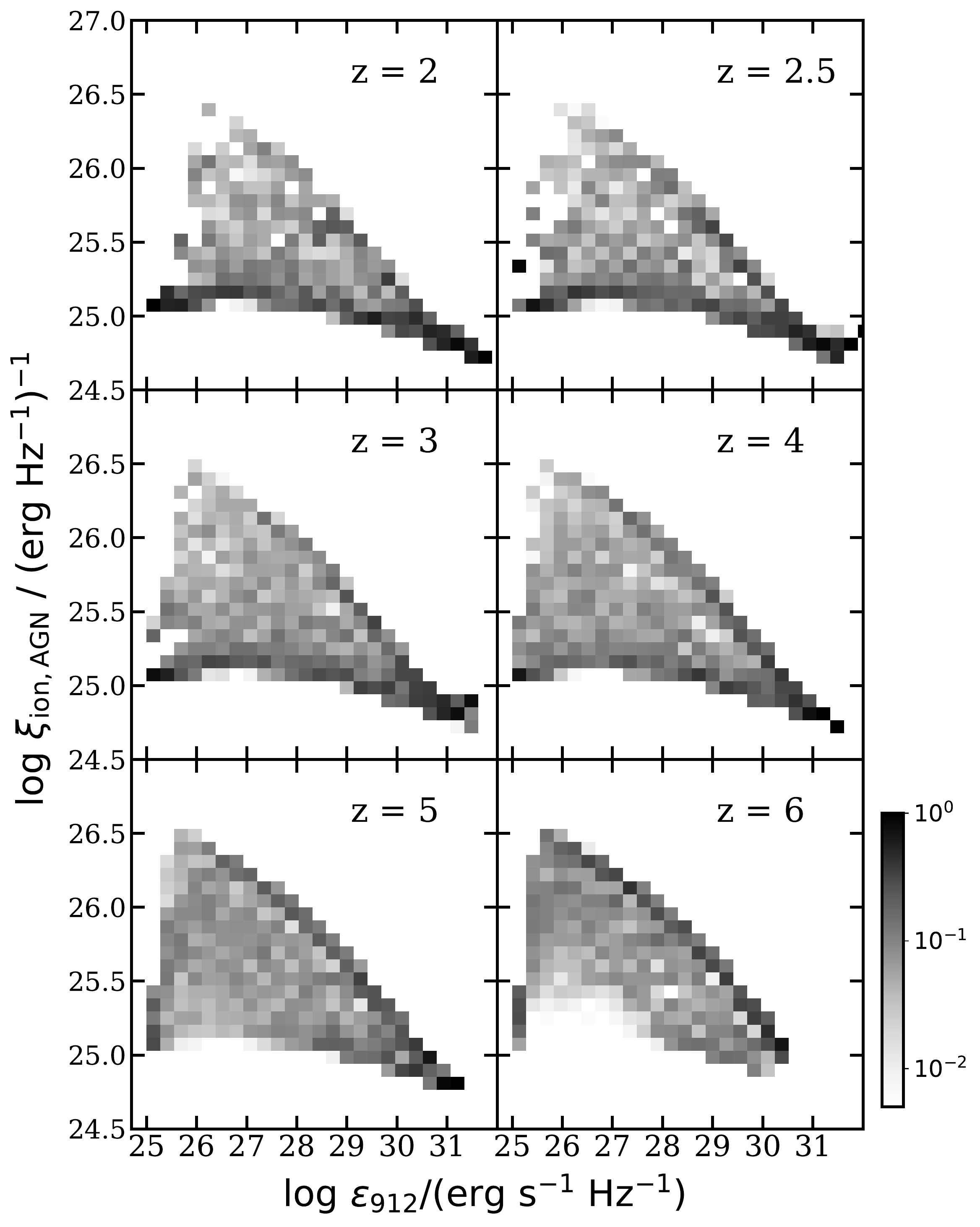}
    \caption{Conditional distributions of $\xi_\text{ion,AGN}$ versus $\epsilon_{912}$ between $z = 2$ to 6, predicted by our fiducial model. The 2D histogram is colour-coded with the conditional number density per Mpc$^3$ of AGN in each bin, normalized to the sum of the number density in the corresponding (vertical) $\epsilon_{912}$ bin.}
    \label{fig:eps912_nHe_6panels}
\end{figure}

In Fig.~\ref{fig:eps912_nHe}, we show the distribution between $\epsilon_{912}$ and \nionHeII\ and between $\epsilon_{912}$ and the AGN \ion{He}{II} ionizing production efficiency, defined as $\xi_\text{ion,AGN} \equiv \dot{N}_\text{HeII}/\epsilon_\text{912}$, for the predicted AGN populations at $z=3$. The physical quantities shown in this plot and the model components related to these predictions do not do not have any explicit redshift dependence.
For both relations, the top and bottom boundaries are marked by the Eddington accretion rate and the cut-off accretion rate where AGN become radiatively inefficient. Both panels show that the scatter in these relations arises mainly from the BH accretion rate, which has a strong effect on the shape of the AGN spectra (see top panel in Fig.~\ref{fig:QSOSED_diagnostics}). We also predict that this scatter increases rapidly towards the UV-faint AGN populations, where the scatter in $\xi_\text{ion,AGN}$ can be over a dex for AGN with $\epsilon_{912} \lesssim 28$. Therefore, $\epsilon_{912}$ may not be an effective tracer for the hard ionizing photon production rate. 
Fig.~\ref{fig:eps912_nHe_6panels} is a similar figure where we show the conditional distributions of $\xi_\text{ion,AGN}$ versus $\epsilon_{912}$. These 2D histograms are colour-coded to reflect the number density weighted distribution in each of the corresponding (vertical) $\epsilon_{912}$ bins among AGN populations. Furthermore, while Fig.~\ref{fig:eps912_nHe} shows that there is a large scatter in the $\epsilon_{912}$--$\xi_\text{ion}$ relation, the number density weighted distributions shown in Fig.~\ref{fig:eps912_nHe_6panels} break down the contribution from AGN across a range of redshifts. When accounting for the evolution between AGN number densities and their accretion rate (similar to that of Fig.~\ref{fig:BH_diagnostics}), we see that $\epsilon_{912}$ is even less reliable as a tracer for the collective contribution from all AGN (e.g. cosmic ionizing emissivity \nionHeII) given the redshift evolution in BH accretion rates across the AGN populations.

In our estimates of these \textit{photon budgets}, we adopt the photon production rate contributed by all predicted physical sources, without accounting for the effects from dust and gas obscuration. 
Instead, these effects will be folded into the overall ionizing photon escape fraction. 
This is a modelling decision we made with the intent to keep a clean separation between the \textit{production} of ionizing photons by the full population of sources, the \textit{fraction} of ionizing photons that were able to escape to the IGM, and whether or not a given AGN is \textit{detected} at a specific wavelength with a specific detection limit. 
We emphasize that this is different from the conventional \textit{emissivities} reported by other studies, which only account for the \textit{detected} AGN populations at their observed luminosities. 
For instance, \citet{Hopkins2007}, \citet{Haardt2012}, \citet{Kulkarni2019}, and \citetalias{Shen2020}, estimate the 1450\AA\ emissivity ($\epsilon_{1450}$) by integrating observationally fitted AGN UVLFs over a certain luminosity range (see Fig.~\ref{fig:Compare_UV1450_dust_LFs}), where extinction and obscuration are implicitly accounted for. This is then converted to 912\AA\ emissivity assuming a power-law spectrum.
On the other hand, \citet{Madau2015} provides an empirical estimate that is motivated by the set of observational constraints that are shown in Fig.~\ref{fig:AGN_emissivity}, which are also derived from a set of FUV \citep{Bongiorno2007,  Glikman2011, Masters2012, Giallongo2015} and optical observations \citep{Bongiorno2007, Palanque-Delabrouille2013}, where similar extrapolation and power-law assumptions were applied.
In effect, \textit{these previous studies assume that AGN that are not detected at 1450\AA\ do not contribute anything to the ionizing photon budget}. Yet, we have shown that torus-scale obscuration, which is expected to be highly anisotropic, can easily decrease the normalization of the observed 1450\AA\ LF by an order of magnitude or more. Although an AGN with a torus viewed at a small inclination angle may be so obscured along our line of sight that it drops out of a UV-selected survey, it is expected that photons, including ionizing photons, should be able to escape along directions within some solid angle perpendicular to the torus, thus potentially contributing to the global budget of ionizing photons.
Indeed, there is strong observational evidence that ionizing radiation from  quasars can be highly anisotropic \citep[e.g.][]{Lau2017}.
This difference largely explains why our predictions are approximately an order of magnitude higher than previous predictions from the literature, which were all based on observed UV 1450\AA\ luminosity functions. In addition, it can be seen from Fig.~\ref{fig:Compare_UV1450_dust_LFs} and Fig.~\ref{fig:AGN_Bol_LFs} that the predictions of our physics-based models tend to yield steeper faint end slopes for the AGN LFs than the assumed values that are extrapolated in previous studies. Finally, as we have shown, our physically based AGN spectral synthesis model yields somewhat different results than the commonly assumed fixed power law spectrum or template SED.
To illustrate the impact of adopting conventional conversions assuming power-law spectra and simplified scaling relations, we show in Fig.~\ref{fig:AGN_emissivity}, an alternative comoving emissivity $\epsilon_{912}$  calculated assuming a scaling relation between \mBH\ to $B$-band luminosity and a power-law conversion to $\epsilon_{912}$ similar to the one used in \citet[][also see the orange dashed line in Fig.~\ref{fig:eps912_compare}]{Hassan2018}. We find that this yields an overestimate in the comoving emissivity of about a factor of two relative to our full physics-based model predictions.

\begin{figure*}
    \includegraphics[width=2\columnwidth]{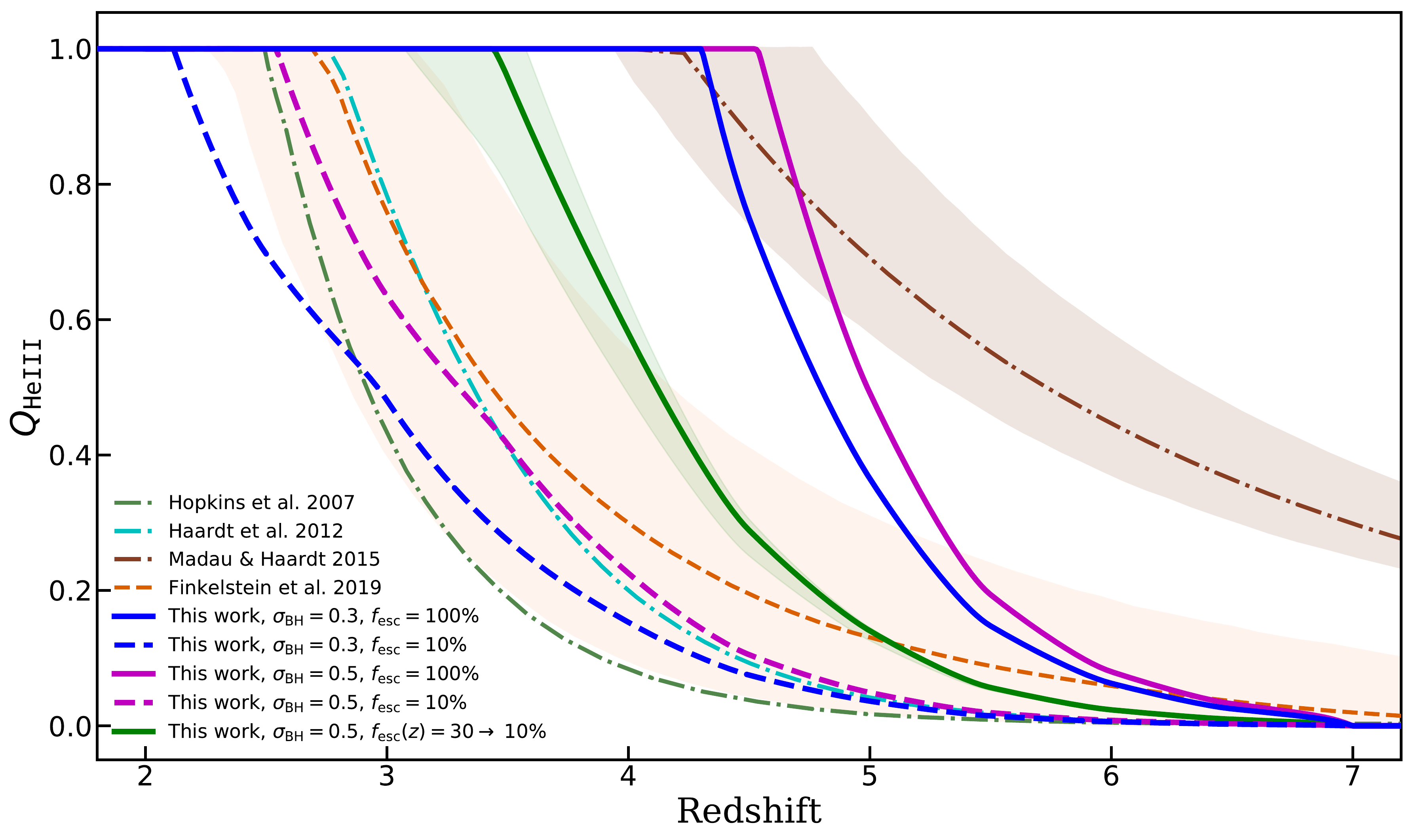}
    \caption{The \ion{He}{III} volume filling fraction, $Q_\text{\ion{He}{III}}$, as a function of redshift. Solid blue and purple lines show predictions from our fiducial model ($\sigma_\text{BH}=0.3$) and adjusted model ($\sigma_\text{BH}=0.5$) with an escape fraction of unity, and blue and purple dashed lines show the predictions with a constant $f_\text{esc}=0.1$. 
    We also show two additional cases for the adjusted model, where \fesc\ evolves as a function of redshift from 0.30 at $z\gtrsim7$ to 0.10 at $z\sim2$ (green). The shaded regions correspond to a range of evolutionary paths for the escape fraction. See text and Fig.~\ref{fig:fesc_redshift} for details. 
    These cases are colour-matched to the escape fraction evolution plotted in Fig.~\ref{fig:fesc_redshift}.
    Our results are compared with past studies by \citet{Hopkins2007}, \citet{Haardt2012}, \citet{Madau2015}, and \citet{Finkelstein2019}, as shown in the figure label.} 
    \label{fig:QHeII_redshift}
\end{figure*}

\begin{figure}
    \includegraphics[width=\columnwidth]{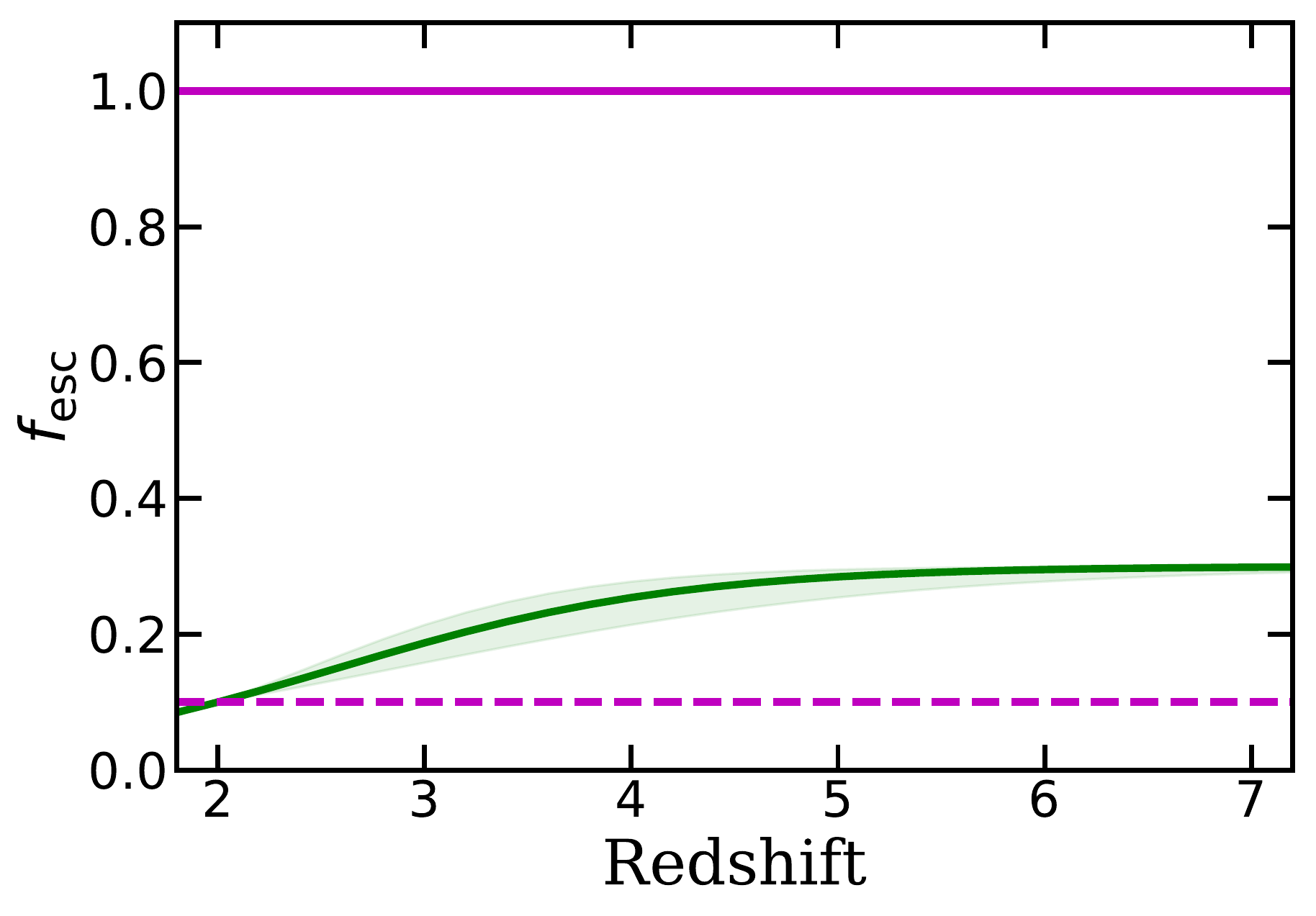}
    \caption{Illustration of how escape fraction evolves with redshift in the various scenarios we consider in this work. 
    The green line shows a scenario where the escape fraction evolves from 0.30 to 0.10 between $z\sim 7$ and 2 according to Eqn.~\ref{eqn:fesc} with $k=1.2$. The top and bottom boundaries of the green shaded region show alternative scenarios with $k=1.6$ and $k=0.8$ to illustrate how sensitive the model output is to the assumed escape fraction.
    The purple solid and dashed lines show a constant escape fraction of $f_\text{esc} = 1.00$ and 0.10.}
    \label{fig:fesc_redshift}
\end{figure}

To obtain the predicted ionizing photon budget, we apply an overall escape fraction that folds in the effects of attenuation and obscuration, which is applied to all AGN when modelling the comoving hard ionizing photon emissivity and the subsequent volume filling fraction of \HeIII. 
As reported by numerical simulations, the escape fraction for star-forming galaxies can be very stochastic depending on the internal structures and the many intricate physical processes occurring in individual halos and even individual  star forming regions \citep{Paardekooper2015, Xu2016, Ma2020} and some similarities can be drawn for active galaxies.
We add that the escape fraction of He ionizing photon from AGN is a highly complex, unsolved problem that has been greatly understudied.
Given the large uncertainties and stochasticity of this quantity both across AGN populations and across time, we adopt a simple empirical approach similar to the one adopted in \citetalias{Yung2020a} and treat the escape fraction as a population averaged quantity, which is the escape fraction of all hard ionizing photons collectively produced by all AGN. 
In this work, we allow \fesc\ to be a constant value or to vary as a function of redshift.
We are in the process of developing a more advanced escape fraction model that accounts for the physical properties of individual sources (Yung et al. in preparation).

In this work, we consider two bracketing cases where we allow a maximum escape fraction of \fesc\ = 100\% and a minimum escape fraction of \fesc\ = 10\%. 
In addition, we consider a redshift-dependent average escape fraction modelled with the logistic equation as presented in \citetalias{Yung2020a}:
\begin{equation}
    f_\text{esc}(z) = \frac{f_\text{esc,max}}{1+\left(\frac{f_\text{esc,max}}{f_\text{esc,0}}-1\right)\;e^{-k_0(z-z_0)}}\text{,}
    \label{eqn:fesc}
\end{equation}
assuming \fesc\ decreases from \fesc$_\text{,max}$ at high redshift at a characteristic rate $k_0$ until it asymptotically reaches \fesc$_\text{,min}$ at an anchoring redshift $z_0$. 
This can be loosely interpreted as a \emph{population-averaged} quantity, where all AGN share the same value, or as the effective escape fraction of the total number of ionizing photon collectively produced by all AGN.
This approach allows us to roughly estimate the plausible range for the global escape fraction in order to reproduce observational constraints.

Based on this parametrization, we explore a scenario which the \textit{population-averaged} escape fraction evolves from a minimum value \fesc$_\text{,min} = 0.1$ at $z_0 = 2$ to \fesc$_\text{,max} = 0.30$ at $k_0 = 1.2 \pm 0.4$. This is illustrated in Fig.~\ref{fig:fesc_redshift}, along with the two other non-evolving cases of \fesc\ = 1.0 and 0.1 considered in this section. Here, \fesc$_\text{,max} = 0.30$ is chosen very loosely to represent a more realistic approximation of ionizing photons are escaping averaged across all AGN. 
We note that the value of $f_\text{esc,max}$ adopted here is in rough agreement with the fraction of unobscured AGN at $z \sim 2$ reported in \citet{Buchner2015}. 
The growth rate $k_0 = 1.2$ is empirically set by hand to reproduce a relatively smooth transition from \fesc$_\text{,min}$ to \fesc$_\text{,max}$. The upper bound of the shaded regions in Fig.~\ref{fig:fesc_redshift} corresponds to $k_0 = 1.6$, where \fesc\ remain at \fesc$_\text{,max}$ longer before dropping down to \fesc$_\text{,min}$. Conversely, the lower bound corresponds to $k_0 = 0.8$, which \fesc\ at a more rapid pace.
We also note that while the parametrization for AGN \fesc\ is the same as the one we adopt for star-forming galaxies in \citetalias{Yung2020a}, the parameters for the escape fraction for AGN populations are set independent of our previous work.

In Fig.~\ref{fig:QHeII_redshift}, we show predictions for the redshift evolution of the \textit{volume-averaged} IGM \HeIII\ volume filling fraction, $Q_\text{\HeIII}$ obtained by solving equation (\ref{eqn:He_reion}). 
We show the two bracketing cases of $f_\text{esc,He} = 100\%$ and $10\%$. 
While the former is unlikely given that a large fraction of galaxies are known to be obscured, this gives an estimate of the absolute upper limit of how quickly helium could have been reionized given the total amount of ionizing photons produced. On the other hand, adopting \fesc $= 10\%$ yields results that are generally in good agreement with estimates reported by past studies, and are consistent with  observational constraints indicating that \HeII\ reionization is likely still in progress at $z\sim2.7$ \citep{McQuinn2009, Shull2010, Syphers2014}. 
In addition, \ion{H}{I} Ly$\alpha$ forest measurements indicate a bump in the IGM thermal history at $z \sim 2.8$, which has been interpreted as an indirect indicator for the end of \HeII\ reionization \citep{Schaye2000, Becker2011, Becker2013, Puchwein2015, UptonSanderbeck2016, Hiss2018}. We note that these broad indicators do not constrain the early stages of the phase transition nor the variance due to the overall patchiness of the process.
We find that either of our redshift-dependent $f_\text{esc}$ scenarios are capable of reionizing the IGM well before $z\sim4$. 
The tabulated data for the predicted ionizing emissivity and reionization history are summarized in Table~\ref{table:nionHeII}. 

\begin{table}
    \centering
    \caption{Tabulated data for the \ion{He}{II} ionizing emissivity \nionHeII\ (s$^{-1}$ Mpc$^{-3}$), specific emissivity $\epsilon_{912}$ (erg s$^{-1}$ Mpc$^{-3}$ Hz$^{-1}$), and reionization history between $z=2$ to 7, assuming an ionizing photon escape fraction of 100\% and 10\%. }
    \label{table:nionHeII}
    \begin{tabular}{cccccc}
        \hline
        & & & \multicolumn{2}{c}{$Q_\text{\ion{He}{III}}$} & \\
        & & & \multicolumn{2}{c}{$\sigma_\text{BH} = 0.5$} & \\
        redshift & $\log$ $\epsilon_{912}$ & $\log$ \nionHeII & \fesc=100\% & \fesc=10\%  \\
        \hline
        2.0 & 26.21 & 51.18 & 1.00 & 1.00 \\
        2.5 & 26.24 & 51.23 & 1.00 & 1.00 \\
        3.0 & 26.08 & 51.17 & 1.00 & 0.63 \\
        3.5 & 26.00 & 51.13 & 1.00 & 0.42 \\
        4.0 & 25.80 & 50.99 & 1.00 & 0.23 \\
        4.5 & 25.45 & 50.76 & 1.00 & 0.11 \\
        5.0 & 25.10 & 50.53 & 0.49 & 0.05 \\
        5.5 & 24.66 & 50.20 & 0.19 & 0.02 \\
        6.0 & 24.18 & 49.89 & 0.08 & 0.01 \\
        6.5 & 23.71 & 49.56 & 0.03 & 0.00 \\
        7.0 & 23.20 & 49.22 & 0.00 & 0.00 \\
        \hline
    \end{tabular}
\end{table}

We also compare our predictions with past studies that adopt a similar reionization model.
The \citet{Finkelstein2019} model allows a large degree of freedom in the AGN comoving 1-ryd specific emissivity spanning a range between the UVLF-based result from \citet{Hopkins2007} and the AGN-dominated result from \citet{Madau2015} (see Fig.~\ref{fig:AGN_emissivity}). The AGN emissivity and subsequent reionization history, along with the high-redshift galaxy UVLFs and the Lyman-continuum production efficiency, are then jointly constrained by a collection of observational constraints using a Markov Chain Monte Carlo (MCMC) approach. 
The orange line and shaded region show the median and the 68\% central confidence range from their fiducial model for comparison.
The line labelled \citealt{Hopkins2007} is a special case from the \citep{Finkelstein2019} model where the AGN 1 ryd emissivity is restricted to follow the results of \citet{Hopkins2007}. 
We note that our model configurations (IGM mean comoving helium number density and recombination time-scale) are consistent with the assumptions in \citet{Finkelstein2019}.
The \citet{Haardt2012} analytic model also adopts an AGN $\epsilon_{912}$ that is in close agreement with the results of \citep{Hopkins2007}, which provides a good example of how even with the same assumed underlying AGN population, modelling assumptions adopted by different studies, such as the conversion to \nionHeII\ and escape fraction, may impact the final predicted reionization history.
The \citet{Madau2015} model presents an extreme scenario where AGN completely reionize both intergalactic hydrogen and helium without any contribution from galaxies. We show results from their default model configurations, as well as the range corresponding to changes in clumping factor, IGM temperature, and EUV spectral slope. We find that even our most `optimistic' model with  $f_\text{esc, He}=1$ does not produce the very early helium reionization predicted by the \citet{Madau2015} model. Our fiducial and adjusted models with $f_\text{esc, He}=0.1$ produce predictions for the helium reionization history that are within the range of previous studies such as  \citet{Hopkins2007}, \citet{Haardt2012}, and \citet{Finkelstein2019}.

\begin{figure}
    \includegraphics[width=\columnwidth]{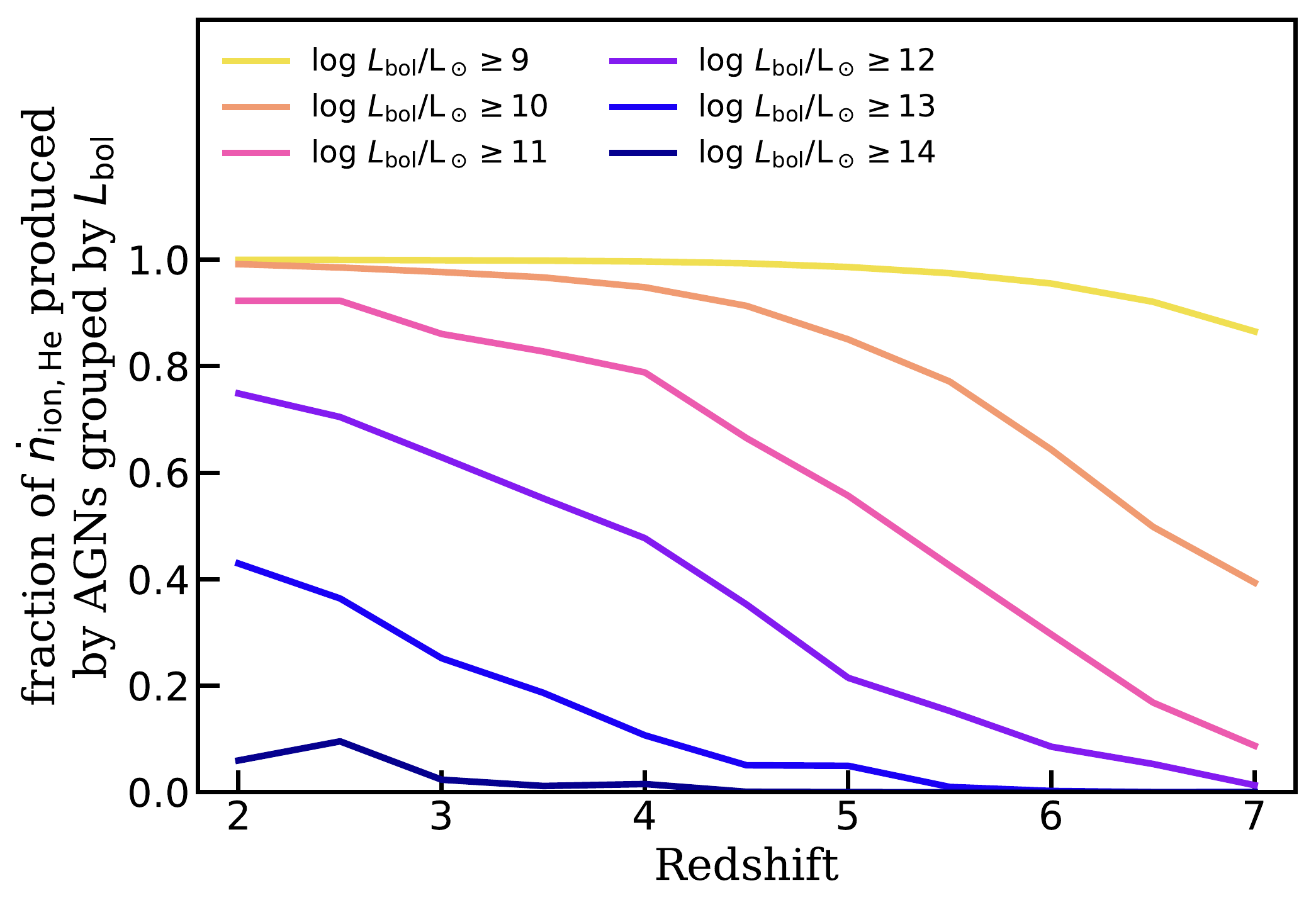}
    \caption{Predicted redshift evolution for the cumulative fraction of \ion{He}{II} ionizing photons contributed by AGN above various bolometric luminosity limits.
        These predictions show the intrinsic production rate and do not account for the escape fraction that possibly varies across the AGN population. These estimates can be used to assess the completeness of ionizing photon estimates in surveys where faint AGN are not fully sampled. }
    \label{fig:grouped_nionHefrac_agn_Lbol}
\end{figure}

\begin{figure}
    \includegraphics[width=\columnwidth]{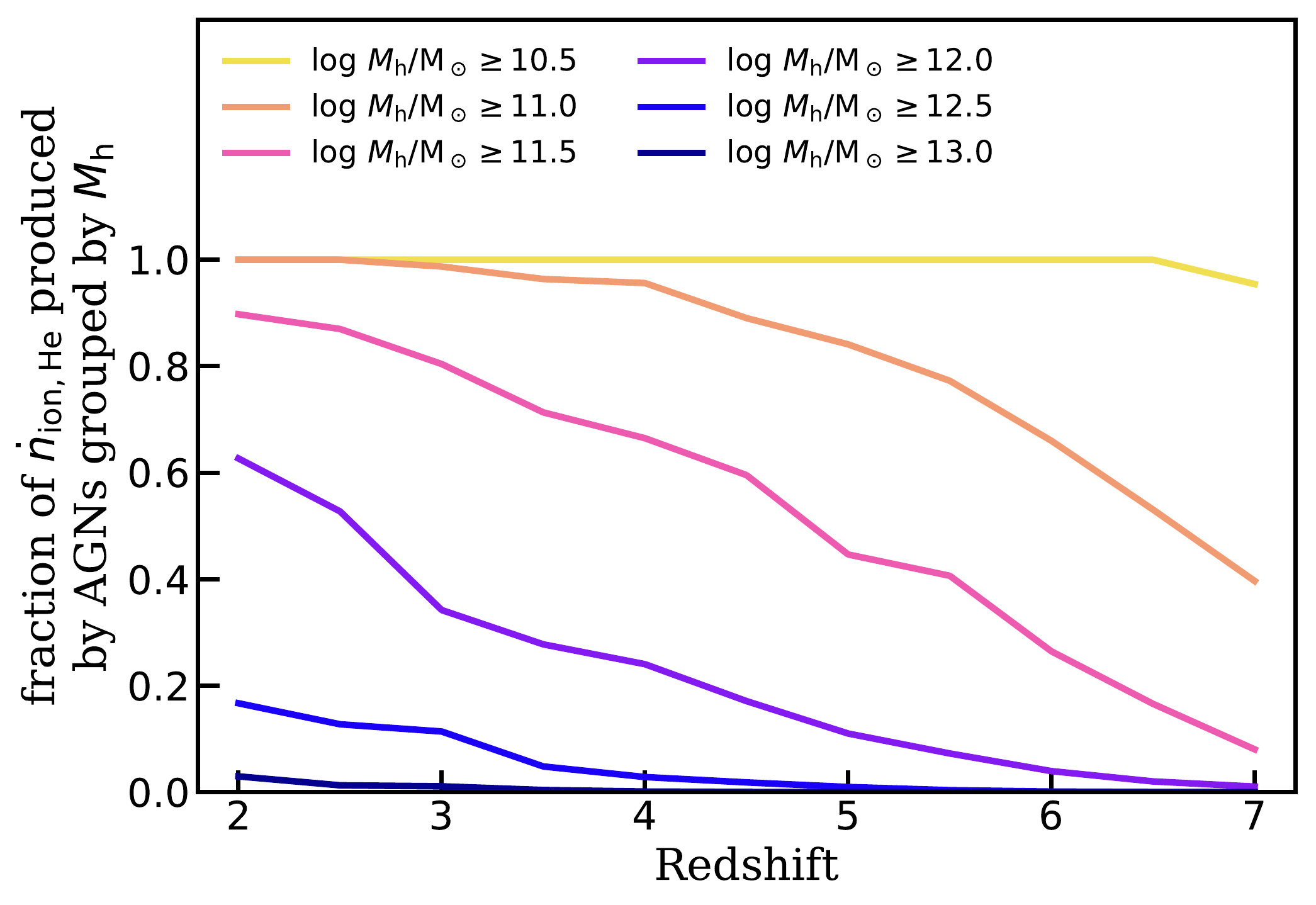}
    \caption{Predicted redshift evolution for the cumulative fraction of \ion{He}{II} ionizing photons contributed by AGN hosted in halos above various halo mass limits.
        These predictions reflect the intrinsic production rate and do not account for the escape fraction, which may depend indirectly on halo mass. This can be used to assess the completeness of ionizing photon estimates in simulations where the full dark matter halo population is not fully sampled due to mass resolution limits or limited simulation volume. }
    \label{fig:grouped_nionHefrac_agn}
\end{figure}

\begin{figure}
    \includegraphics[width=\columnwidth]{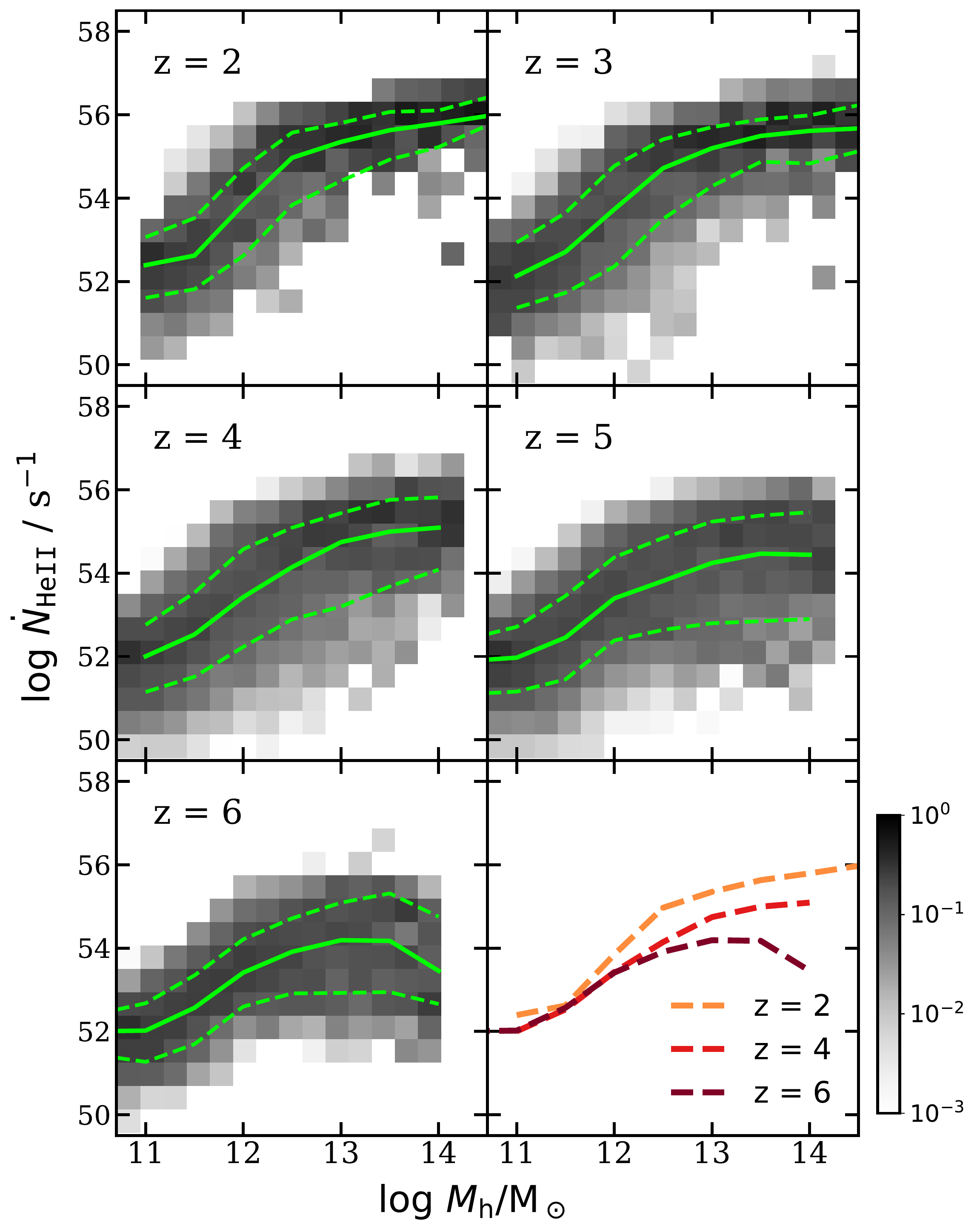}
    \caption{The production rate of \ion{He}{II} ionizing photons, \NionHeII, as a function of AGN host halo mass between $z = 2$ to 6, predicted by our fiducial model. The 2D histogram is colour-coded with the conditional number density per Mpc$^3$ of AGN in each bin, normalized to the sum of the number density in the corresponding (vertical) halo mass bin. The green solid and dashed lines mark the 50th, 16th, and 84th percentiles. The last panel shows an overlay of the median \NionHeII\ values at $z = 2$, 4, and 6 from our model.}
    \label{fig:NHe_mhalo_z}
\end{figure}

An advantage of our physics based model, which predicts the ionizing photon rate for each galaxy and halo, is that we are able to further examine the contribution of ionizing photons from AGN of different halo masses and bolometric luminosities. 
In Fig.~\ref{fig:grouped_nionHefrac_agn_Lbol}, we break down the contribution from AGN by bolometric luminosity between $z = 2$ to 7. 
This intuitively illustrates which AGN are dominating the production of hard ionizing photons at any given time.
As expected, at earlier times the majority of ionizing photons are dominated by the more prevalent faint AGN, which then gradually transition to the brighter AGN as they grow in number density.
These results can also be used to estimate the completeness of ionizing sources captured by a given survey. 
For example, luminous AGN above the detection limit for a typical \textit{JWST} wide-field survey at $\log(L_\text{bol}/L_\odot) \sim 11$, as shown in Fig.~\ref{fig:AGN_Bol_LFs}, are collectively producing $\sim 40\%$ of the total ionizing photons. Note that these cuts are made based on bolometric luminosity and do not account for the survey volume, which is a separate requirement for properly capturing the very rare high-redshift quasars.

Similarly, Fig.~\ref{fig:grouped_nionHefrac_agn} shows a breakdown for the contribution of He ionizing photons as a function of the host halo masses. This is helpful for assessing the fraction of ionizing photons that are captured in numerical simulations, where low-mass halos may fall below the mass resolution.

Furthermore, we investigate how the production rate of hard ionizing photons scales with the AGN host halo mass. 
In Fig.~\ref{fig:NHe_mhalo_z}, we present the predicted scaling relation and distribution for \HeII\ ionizing photon production rate, \NionHeII, with respect to the host halo mass between $z=2$ to 6. 
Tabulated data for the median and scatter are given in Table~\ref{table:NHe_mhalo}.
We find that the redshift evolution in this relation is driven by a subtle increase in both $\dot{M}_\text{acc}$ and \mBH\ in halos of similar mass towards lower redshifts, which both contribute to a higher ionizing photon count for a halo at a given mass. 
We note that the most massive halos presented in this work are extremely rare, and their contributions to the overall ionizing photon budget are extremely low as illustrated in Fig.~\ref{fig:grouped_nionHefrac_agn}.
We also note that Fig.~\ref{fig:NHe_mhalo_z} only includes radiatively efficient AGN ($\log\ \dot{m} \gtrsim -1.7$) and their host halos. 
Halos with a radiatively inefficient AGN or no AGN are not included.
These results can be adopted in (semi-)numerical simulations that assume simple scaling relations between ionizing photon production rate and stellar mass or halo mass \citep[e.g.][]{Hassan2016, Hassan2018}. 
This result is analogous to one of the results presented in \citetalias{Yung2020}, where we explored the ionizing photon production rate by galaxies as a function of halo mass.

\section{AGN contribution to cosmic hydrogen reionization}
\label{sec:H_reion}
In \citetalias{Yung2020} and \citetalias{Yung2020a}, we presented predictions from a detailed physical model showing estimates for the production of ionizing photons from high-redshift galaxies and the implied reionization history of intergalactic hydrogen.
The results indicate that these galaxies, assuming a Lyman-continuum (LyC) escape fraction that is broadly consistent with other studies, are likely to have produced sufficient ionizing photons to fully reionize the Universe within the time-frame consistent with the observed IGM and CMB constraints. In these studies, the contribution from AGN at $z\gtrsim 4$ were assumed to be negligible.
In this section, we provide estimates for the contribution to hydrogen reionization from AGN based on the new inclusion of AGN in our modelling infrastructure.

\begin{figure}
    \includegraphics[width=\columnwidth]{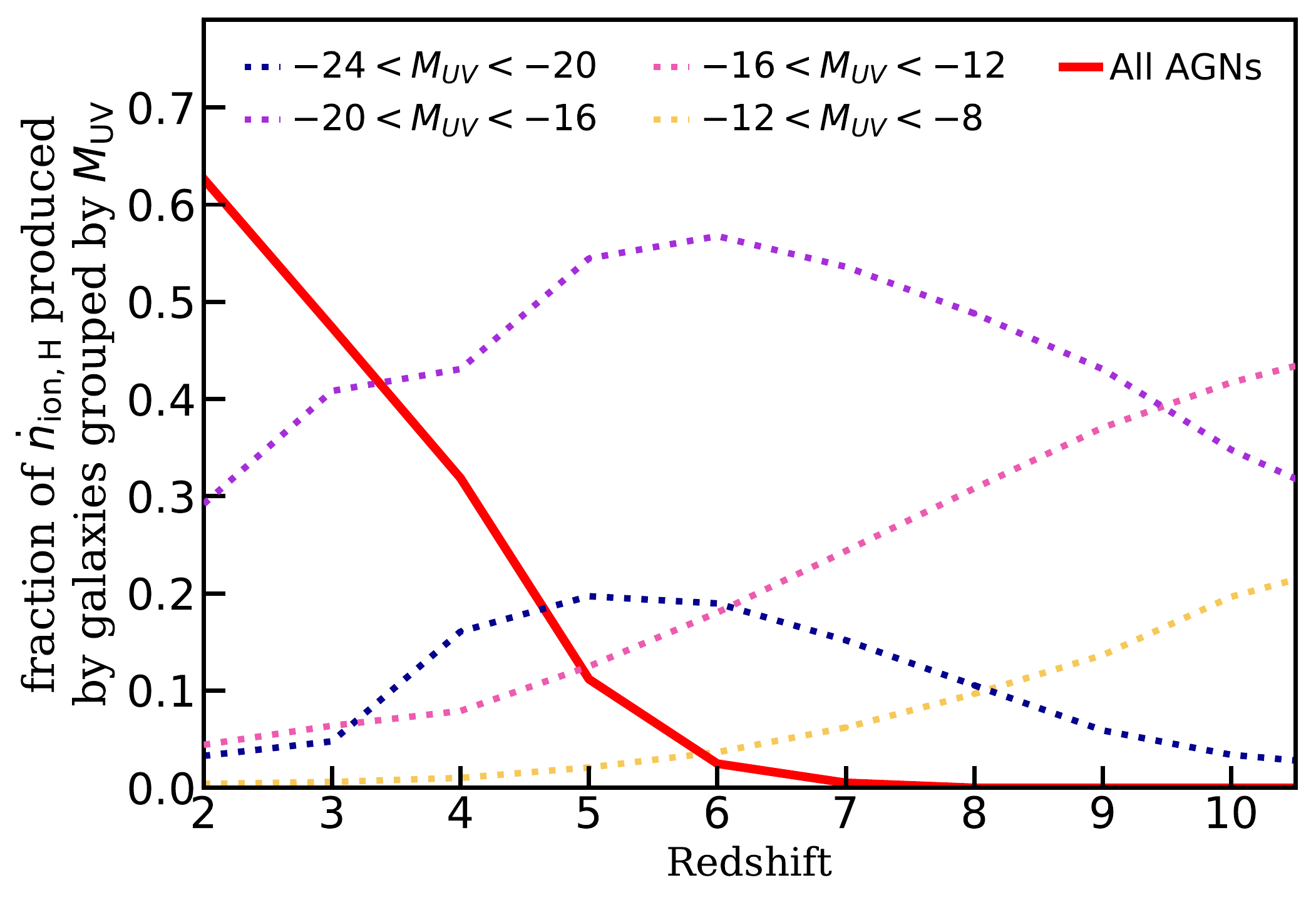}
    \caption{Predicted fraction of cosmic hydrogen ionizing emissivity, \nionH, contributed by AGN between $z = 4$ -- 7 (red solid line) relative to the combined number of ionizing photons produced by AGN and all star-forming galaxies.
    Similar to fig.~12 presented in \citetalias{Yung2020a}, the contributions from star-forming galaxies are divided into groups by their rest-frame dust-attenuated $M_\text{UV}$ (dotted lines).
    Galaxies outside the range of $M_\text{UV}$ shown contribute $<1$ per cent of ionizing photons at all times. }
    \label{fig:grouped_nionfrac_agn}
\end{figure}

The hydrogen ionizing photon production rate, $\dot{N}_\text{H}$, for individual AGN is obtained by integrating equation (\ref{eqn:NionHeII}) between 912\AA\ and 228\AA, or equivalently from 13.6 to 54.4 eV, and the overall contribution from all AGN is calculated using equation (\ref{eqn:nionHeII}), with the helium ionizing photon production rate replaced by $\dot{N}_\text{H}$. 
In order to assess the contribution from AGN at low redshift, we have extended the predictions from previous studies down to $z = 2$ based on the same model configurations and calibrated parameters as presented in \citetalias{Yung2019} and \citetalias{Yung2020a}. In Fig.~\ref{fig:grouped_nionfrac_agn}, we show the fraction of H ionizing photons produced by AGN relative to the combined contributions from AGN and galaxies as a function of redshift. 
We also compare this to the contribution from galaxy populations broken down by rest-frame UV luminosity, which can be directly compared to fig.~12 in \citetalias{Yung2020a}. We see that the contribution from AGN during the epoch of hydrogen reionization,  $6 \lesssim z \lesssim 10$, is fairly negligible, but the contribution rapidly rises to $\sim 30$ per cent at $z\sim4$ and can even overtake that from all galaxies at $z \lesssim 3$. We note that this plot shows the \emph{intrinsic} production rate of ionizing photons, and does not account for the escape fraction, which may differ between AGN and galaxies, and may vary as a function of halo mass and other galaxy properties.

\begin{figure}
    \includegraphics[width=\columnwidth]{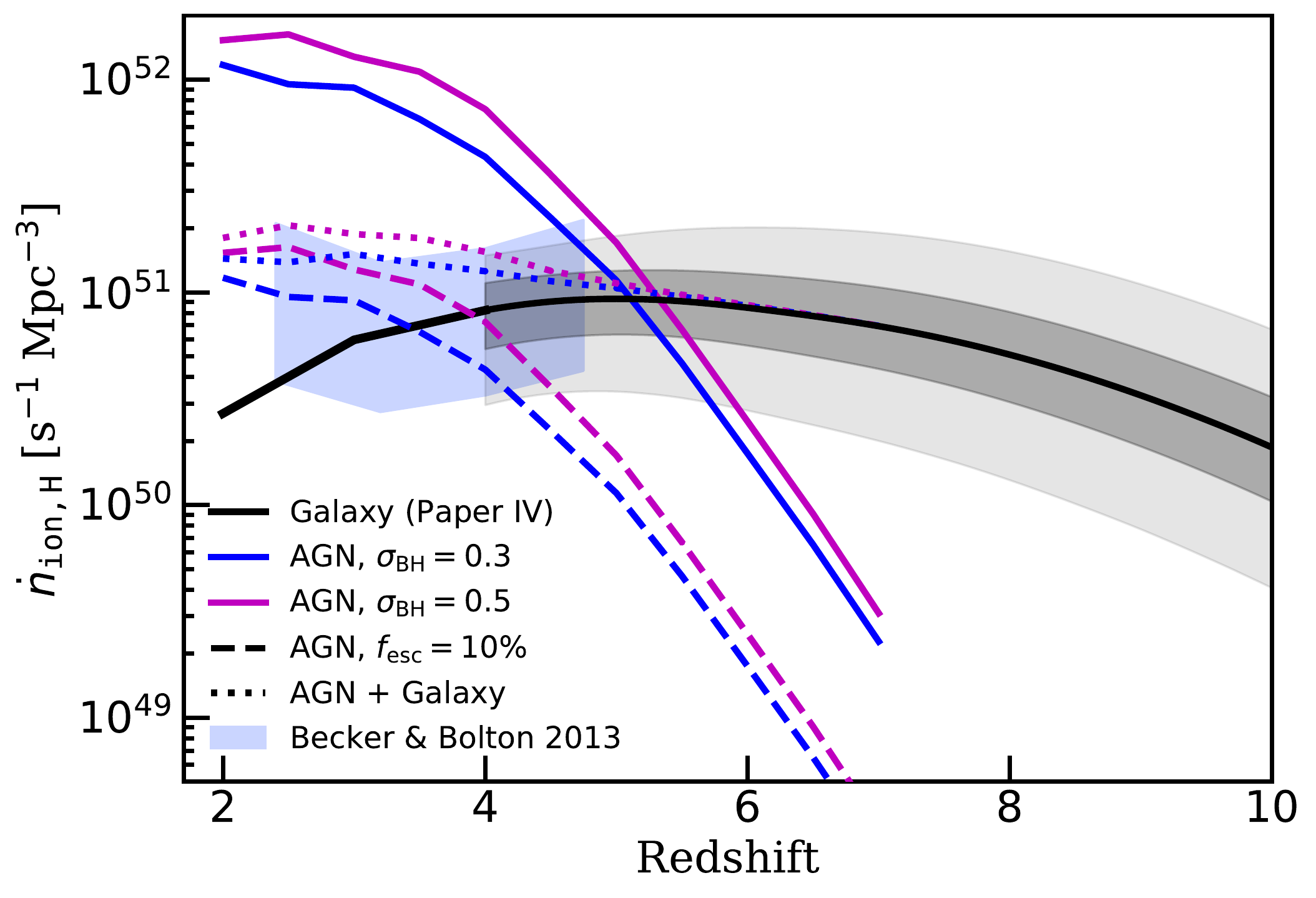}
    \caption{Hydrogen ionizing emissivity contributed by AGN and galaxies as a function of redshift. The purple (blue) line shows predictions from our adjusted (fiducial) model, where the solid and dashed lines assumes AGN escape fraction of \fesc=1.0 and \fesc=0.1, respectively. 
    The galaxy contribution includes a redshift-dependent, \textit{population-averaged} \fescstar$(z) \sim 25\% \rightarrow 5\%$ for star-forming galaxies from $z\sim15\rightarrow4$, which was found to reproduce a set of IGM and CMB observational constraints on the reionization history (see section 3.2 and fig.~8 in \citetalias{Yung2020a} for a full discussion).
    These results are compared to observational constraints from \citet{Becker2013}.}
    \label{fig:AGN_nH_emissivity}
\end{figure}

Furthermore, we show the comoving hydrogen ionizing emissivity from AGN as a function of redshift in Fig.~\ref{fig:AGN_nH_emissivity}, assuming a fixed $f_\text{esc} = 1.00$ and 0.10 for AGN. To compare this with the contribution from galaxies, we include one of the key results from \citetalias{Yung2020a}, where a Markov Chain Monte Carlo (MCMC) pipeline was implemented to determine the redshift-evolution of escape fraction for star-forming galaxies, \fescstar, to match the observed \Lya\ constraints from \citet{Becker2013} and the Thomson scattering optical depth of the CMB, \tauCMB, reported by \citet[XLVII][]{Planck2016a}. In this previous study, we reported that a LyC escape fraction that evolves from \fescstar$\sim 25$ per cent at $z \gtrsim 12$ to a few per cent towards the conclusion of reionization at $z \sim 6$ results in good agreement with all available constraints on the reionization history.
See section 3.2 in \citetalias{Yung2020a} for more details.

Although our models predict that it is unlikely that AGN have contributed significantly to the initial reionization of intergalactic hydrogen, we note that their contribution increases more rapidly than that of galaxies and could be comparable to that of galaxies at around $z \sim 4$. 
If this additional AGN contribution were accounted for in the MCMC framework as in \citetalias{Yung2020a}, it would likely result in even lower values for the implied escape fraction from galaxies at these redshifts (depending, of course, on the corresponding escape fraction from AGN). With fairly plausible assumptions for the escape fraction, the combined contribution from AGN and galaxies is consistent with the \citeauthor{Becker2013} constraints at $z \sim 2$-4. The fairly steep rise in emissivity due to the AGN contribution may provide an explanation for the bump at $z \sim 3$. Furthermore, AGN could serve as an `insurance policy' in case the escape fractions in galaxies are found to be lower than those favored in our models. In this case, the contribution from AGN could help complete hydrogen reionization.

Untangling the degenerate contribution from AGN and galaxies towards the end of cosmic hydrogen reionization has proven to be very difficult with the current set of constraints on IGM thermal history, given that each class of objects has their own set of uncertainties in number density, ionizing production rate,  and ionizing photon escape fraction. 
In this work and the rest of the series, we have demonstrated that a detailed physically-based source model that self-consistently models the co-evolution of galaxies and their black holes can provide a novel approach to gaining insights into this puzzle. 
Future work on more physically grounded modelling of ionizing photon escape fraction together with anticipated constraints on the faint galaxy nad AGN population that will be obtained by \textit{JWST} will bring further insights that will help untangle the contribution from AGN and galaxies.

\section{Discussion}
\label{sec:discussion}

This \textit{Semi-analytic forecasts for JWST} paper series aims to provide a comprehensive collection of predictions for observable properties and underlying physical properties of objects forming in the early universe. 
Predictions from these successful models presented earlier in the series were used to make detailed predictions for the expected properties of galaxies that have been used to facilitate the planning of upcoming surveys, as well as in the development of data reduction and interpretation pipelines. 
This latest addition to the work series expands our framework to include high-redshift AGN populations and establish the connections between the predicted \textit{intrinsic} AGN properties and their spectral properties, as well as the impact on the reionization history.

The semi-analytic approach represents a `middle way' between empirical or semi-empirical models and numerical simulations. 
This approach uses a selection of models and recipes to track the physical processes that occur at vastly different physical scales. This physically motivated approach preserves the physical origins behind the predicted phenomenological relations and enables us to connect these relations to the underlying physics. 
This approach also minimizes tracking of spatial elements, such as velocity and position, making it very efficient at sampling objects across a wide mass range at a fraction of the computational cost. 
An advantage of this approach over purely empirical models is that it is based on self-consistent underlying co-evolution of black holes and their galactic hosts according to physical laws. 
In addition, the highly flexible, modular structure of our modelling pipeline enables free exploration of different model configurations, including the choice of physical parameters and the recipe or prescriptions for specific physical processes.

\subsection{High-redshift AGN populations and their role in cosmic reionization}
The nature of the sources of reionization is one of the fundamental questions in cosmology. The interplay between AGN, which have harder spectra but are more rare, and galaxies, which are less efficient at producing ionizing photons, but are more common, is one of the primary open questions. Direct observational constraints on the faint AGN population at high redshift are not currently available. Therefore, simultaneously considering the observational constraints on hydrogen and helium reionization provides a promising avenue to address this issue. In the past, this has usually been addressed using empirical descriptions of the AGN luminosity function (extrapolated below the observed population) combined with simple scaling relations or template SEDs. In this work, we present a significant step forward by using a physics-based model, which we confirm produces AGN LFs that are consistent with observations within the uncertainties. This model is coupled with the physical AGN spectral synthesis model of \citetalias{Kubota2018}, which we showed predicts considerably more scatter in key quantities than is generally reflected in the simplified scaling relations. Moreover, as these empirical approaches are typically based on bright (observable) AGN, we showed that they may systematically overestimate the ionizing photon budget.

\subsubsection{Helium reionization}
According to our predictions, if the AGN escape fraction were 100\% (as assumed in many previous studies), helium would be ionized as early as $z\sim 5$. As this is inconsistent with observations which suggest that helium reionization is still in progress at $z\sim 2.7$, we suggest that a more likely scenario, and one that is consistent with our state-of-the-art knowledge about the fraction of obscured AGN and multi-wavelength AGN luminosity functions, is that the effective AGN escape fraction for helium ionizing photons is closer to 10\% on average. Of course, how the effective escape fraction varies from one object to another, or effectively changes for the overall population over cosmic time, is extremely uncertain. We consider a scenario in which the escape fraction is as high as 30\% at $z\sim 7$ and declines to 10\% at $z\sim 2$. This results in the completion of helium reionization by $z\sim 3.5$, a bit early compared to some observational constraints.  Our models predict that the bulk of the helium ionizing photons at $z>4$ are from AGN with bolometric luminosities less than $10^{12} L_\odot$, and are hosted by halos with mass less than $10^{11.5} M_\odot$.

\subsubsection{Hydrogen reionization}
The other side of the coin is whether these AGN play a significant role in  hydrogen reionization. To address this question, we consider the galaxy populations that are self-consistently predicted by these same models, along with their production and emission of ionizing photons, which have also been studied in \citetalias{Yung2020a}. We find that during the epoch of hydrogen reionization ($z>6$), in our models AGN produce 1\% or less of the ionizing photon budget relative to that produced by galaxies. Therefore, we conclude that AGN are unlikely to have contributed significantly throughout most of hydrogen reionization. However, the ionizing photon budget contributed by AGN rises rapidly at $z<6$, with AGN contributing 30\% of the ionizing photon budget at $z=4$ and 60\% at $z=2$ (not accounting for any escape fraction). Thus, it appears that while the ionizing emissivity contributed by galaxies may be starting to decline already by $z\sim 4$, the contribution from AGN is rising steeply over this same interval, perhaps helping to account for the relatively constant ionizing emissivity implied by observations.

\subsection{Our results in the context of other model predictions} 
\label{sec:context_other_model}

We compared our predicted AGN luminosity functions with predictions from a compilation of state-of-the-art hydrodynamic simulations from $z=2$--4 (predictions at higher redshifts were not easily available). It is interesting that these simulations produce a very wide range of predicted AGN LFs over this redshift range, with different normalizations in some cases, and in some cases very different faint end slopes. Our predictions are by no means an outlier in this compilation, and in fact are roughly close to the median between the predictions from different hydrodynamic simulations. This highlights how observed AGN LFs at high redshift can help to constrain the physics regulating black hole accretion in current simulations.

We also anchored and cross checked our model outputs against a variety of observational constraints. For instance, the SEDs for AGN accreting at near-Eddington rates reproduces the observed power-law, as well as the correlation between \mBH\ and $\epsilon_{912}$. We incorporate a detailed estimate of nuclear obscuration using the radiation lifted torus model of \citet{Buchner2017} to forward model our predictions for the hard X-ray luminosity function, and show that they agree well with observational constraints. We also compare our predictions with the estimated observed bolometric luminosity function reported by \citet{Shen2020}, which they obtained by correcting for obscuration and attenuation and performing bolometric corrections on a comprehensive set of multi-wavelength AGN LFs, and find good agreement within the considerable uncertainties.

Although our predicted bolometric LFs are in good agreement with the observational constraints presented by \citetalias{Shen2020}, we predict a cosmic emissivity $\epsilon_{912}$ that is a factor of 10 higher than the one they quote over all redshifts (see  Fig.~\ref{fig:AGN_emissivity}).  We have carried out a thorough comparison between our work and that of \citetalias{Shen2020}, and first ruled out the assumed AGN spectra as the main source of difference (see Fig.~\ref{fig:QSOSED_diagnostics}). Instead, the reason is that \citetalias{Shen2020} calculated the emissivity at $1450$\AA\ ($\epsilon_{1450}$) by integrating the observed UVLF and then assumed a power-law spectra to convert that into $\epsilon_{912}$. Effectively, this assumes that obscured AGN, which are not present in UV-selected samples, do not contribute to the ionizing photon budget. We note that the other past studies compared in Fig.~\ref{fig:AGN_emissivity} also adopt a similar approach to track the ionizing emissivity and are rooted in similar UV1450 observations and empirical relations. However, we have pointed out that there is strong observational evidence that AGN obscuration and the escaping ionizing emission is extremely anisotropic. Therefore, although an AGN might drop out of the sample when observed from a particular line of sight, we would expect ionizing photons to potentially be able to escape over some solid angle in other directions. Therefore, this effective escape fraction (which likely depends on how AGN feedback has cleared gas and dust, producing channels where ionizing radiation can escape) is one of the key parameters in understanding the role of AGN in reionization. For this reason, we advocate for a cleaner separation between ionizing photon production rate and escape fraction. We suggest that the production of ionizing photon should first be estimated based on the unobscured, unattenuated luminosity function. This will have significantly different implications for physical quantities such as the growth rate and mass functions of SMBH at high redshift, which we may eventually be able to test through gravitational wave experiments such as LISA.

\subsection{Major caveats and limitations of current models}

In this subsection we discuss the most significant outstanding observational and theoretical uncertainties, including some caveats about our current models.

Similar to that of galaxies, the ionizing emissivity of AGN can be broken down into three moving parts, the abundance of AGN across cosmic time, their intrinsic production efficiency of ionizing photons, and the fraction of ionizing photons that escape to the IGM. The true abundance of AGN, especially the faint AGN that our models predict are the dominant source of ionizing photons in the early universe, are difficult or impossible to observe with current facilities. The deepest and most complete surveys are in the rest-UV, which may miss many objects either because they become too red to be included in colour-selected QSO surveys or because they become too faint to be detected at all. Enormous progress has been made in recent years in building up the X-ray detected AGN population at high redshift, but for the moment only the brightest objects can be detected at $z>4$. 

For the same reasons, the spectral characteristics of faint, high redshift AGN are also poorly constrained. Many studies adopt scaling relations between AGN bolometric luminosity and the luminosity in a specific band (bolometric corrections), which are obtained from nearby, luminous AGN. As we have pointed out, there can be scatter in these relations, which may correlate with the BH physical properties such as mass and accretion rate, and there are currently few observational constraints on the relevant populations. Even physics-based spectral models such as the one from  \citetalias{Kubota2018}, used here, still contain some free parameters, which are currently calibrated to reproduce a handful of well-measured AGN spectra in the nearby universe. These parameters characterize the physical characteristics, such as electron temperature and spectral index for each of the radiating regions, as well as the radius and scale height of the disc region. Although these parameters are not anticipated to evolve as a function of redshift, some of these characteristics may have implicit dependences on the mass and accretion rate of the accreting black hole, and could vary to some degree for AGN of masses that have not been explicitly tested.

The ionizing photon escape fraction is the least constrained among the three moving parts. We are not aware of any direct observational constraints on the escape fraction of helium or hydrogen ionizing photons from AGN. Moreover, it is extremely difficult to obtain accurate predictions from simulations, since this quantity is certainly highly sensitive to the detailed structure of gas and dust over a vast range of scales, from the scales of the accretion disk and torus around the BH, to the galactic-scale outflows that may clear channels through which photons can escape \citep{Benson2013, Seiler2018}. This is a critical but challenging area for future work.

Turning to the theoretical models used in this work, we describe the main caveats and uncertainties.

We have used an Extended Press-Schechter based algorithm to generate merger trees, as this enabled us to probe a much larger dynamic range than would have been possible with $N$-body simulations. However, the EPS formalism is known to be inaccurate at the factor of 2 level, and is not well tested at extreme redshifts. We have previously compared our predictions for galaxy properties with predictions from other numerical models and simulations \citep{Yung2019a, Behroozi2020}, finding order unity agreement, providing confidence that this is not a significant source of error. EPS merger trees also do not correctly reflect the known correlations, found in N-body simulations, between halo properties such as concentration, and mass accretion history, and do not enable predictions of spatial clustering of halos or their progenitors.

One of the major uncertainties in theoretical models of black hole formation is the mechanism or mechanisms by which seed black holes are formed in the early Universe, which has implications both for the masses of halos in which these objects form and their mass function. Scenarios include seeds from remnants of Pop III stars, which would populate nearly every halo down to very low masses ($\sim 10^{5} M_\odot$) with `light' seeds ($\sim 100 M_\odot$), to direct collapse or gravitational runaway scenarios, which would populate only a fraction of more massive halos with more massive seeds ($\sim 10^4-10^5 M_\odot$; for reviews, see \citep{Volonteri2010, Greene2020}). Most current cosmological simulations simply populate halos above a critical mass with a seed BH with a fixed mass. Our algorithm is similar: we populate every `top level' halo with a seed BH with mass $m_{\rm seed}$. A heavy seeding mechanism could help to explain the difficulty of growing very massive black holes in the very early ($z>7$) Universe without invoking super-Eddington accretion, but current theoretical predictions suggest that it is difficult to produce these seeds at a high enough number density to explain these objects \citep{Visbal2018, Visbal2020}. For recent exploration of the impact of different seeding mechanisms on AGN populations at high redshift, see \citet{Ricarte2018}, \citet{Ricarte2018a}.

A second major uncertainty is the intertwined processes that govern black hole accretion and feedback (which is thought to regulate the black hole accretion as well as star formation). Cosmological simulations typically assume a version of the Bondi-Hoyle model for BH accretion, but recent hyper-refinement simulations with direct capture to measure BH accretion have shown that this approximation can be off by many orders of magnitude, and may not even predict the correct dependence on BH mass \citep{Angles-Alcazar2020}. Cosmological simulations also adopt various different phenomenological approximations to represent AGN feedback, often including multiple feedback `modes' associated with rapid (close to Eddington) or slow (very sub-Eddington) accretion onto the BH \citep[see][and references therein for reviews]{Somerville2015a}. The large dispersion in the predicted AGN LFs from different simulations shown here presumably largely reflects the major uncertainties in modeling these processes. In this work, we adopt the model for black hole growth and feedback first presented in \citetalias{Somerville2008}, and extensively tested in the context of predictions for AGN populations in \citet{Hirschmann2012a}. Although this model is based on a generation of hydrodynamic simulations that is no longer state-of-the-art, we argue that our predictions are still robust for this study, both because we have demonstrated good agreement with observed AGN LFs and because our predictions are consistent with those of recent cosmological hydrodynamic simulations. However, more physical modeling of these processes is clearly a high priority for the next generation of models.

Finally, we adopted a simplified analytic model to estimate the volume filling fraction of ionized hydrogen and helium. 
Similar to our discussion on the reionization model in \citetalias{Yung2020a}, this analytic approach does not capture the potential `patchiness' of the progression in the evolution of ionized volume due to the density fluctuations and clustering of sources, as shown by numerous numerical simulations \citep{Trac2007, Hassan2016, Hassan2017, Mutch2016, Hutter2020}. 
We add that while the epochs of hydrogen and helium reionization are treated as separate events in this work, which is a practice that has been justified and adopted in many similar studies, in reality, the presence of early X-ray sources may heat up both both hydrogen and helium in the IGM, giving rise to an additional channel where intergalactic hydrogen can be collisionally ionized by free electrons arising from photoionized \ion{He}{I} \citep{Venkatesan2001}. This effect will be further explored in an upcoming work where we fully couple the AGN--helium reionization framework presented in this work and the galaxy--hydrogen reionization framework presented in \citetalias{Yung2020a} (Yung et al., in preparation).

In addition, it is evident that AGN emission is highly stochastic and anisotropic \citep{Lau2017}, which causes additional complications for detailed radiation transport modeling. 
As the IGM gas density in highly clustered regions is expected to be higher than the universal average, the shorter recombination timescale will cause a slow down as the ionization front progresses through these regions. As it is known that luminous quasars tend to form in more highly clustered regions than low-mass galaxies, this could have important implications for the relative topology of hydrogen vs. helium, which however our current models cannot capture.

\subsection{Outlook for probing faint high-z AGN with \textit{JWST} and beyond}

One of the main science goals of the highly-anticipated \emph{JWST} is to survey the extremely distant universe, and it is expected that \emph{JWST} will detect objects several orders of magnitude below the current limit of \emph{HST}. 
Photometric and grism surveys with the on-board Near-Infrared Camera (NIRCam) and Near-Infrared Imager and Slitless Spectrograph (NIRISS) will detect large numbers of objects at high redshift. While these upcoming surveys are predominantly searching for star-forming galaxies, they will also presumably detect some AGN. Follow-up spectroscopy with the Near-Infrared Spectrograph (NIRSpec) will enable identification of AGN candidates and will provide insights about their spectral characteristics. Particularly promising is the use of nebular emission line ratios in the rest-UV, which have been shown to provide a more robust discriminator between AGN and SF/composite galaxies at high redshift than traditional BPT diagnostics \citep{Feltre2016, Hirschmann2017, Hirschmann2019}. We intend to extend our models in the near future to enable predictions of such emission lines and explore the potential of \emph{JWST} to identify faint, high-redshift AGN in this manner. 
Moreover, we expect the extended line emission predictions will deliver additional observable signatures that enable direct comparison with low-redshift analogs, such as Green Pea galaxies \citep[e.g.][]{Jaskot2019}.

Planned wide-field survey instruments, such as the \emph{Nancy Grace Roman Space Telescope} \citep[][formerly known as the \emph{Wide-Field Infrared Survey Telescope} or \emph{WFIRST}]{Spergel2015} and the \emph{Vera C. Rubin Observatory} \citep[][formerly the \emph{Large Synoptic Survey Telescope} or \emph{LSST}]{LSST2017} will increase survey coverage by a factor of hundreds compared to current generation instruments, which will dramatically increase the ability to find these rare objects. In the near future, we plan to deliver forecasts in mock lightcone format similar to the ones presented by \citet{Somerville2021} but of multiple square degrees (Yung et al., in preparation).

Looking further into the future, upcoming \textit{approved} facilities, such as the X-ray \textit{Advanced Telescope for High ENergy Astrophysics} (\textit{ATHENA}) mission and the gravitational wave detector \textit{Laser Interferometer Space Antenna} (\textit{LISA}), as well as the proposed NASA X-ray mission concepts Advanced X-ray Imaging Satellite (AXIS) and Lynx \citep{Weisskopf2015}, will enable multi-wavelength and multi-messenger surveys that will provide unprecedented constraints on the high redshift population of supermassive black holes and their accretion rates.

\section{Summary and conclusions}
\label{sec:snc}

In this work, we constructed a computationally efficient modelling pipeline that is based on the versatile Santa Cruz SAM. 
Benefiting from the wide dynamic range achievable by the SAM, we are able to self-consistently model a wide range of objects, including the ones that are too rare to be captured in cosmological hydrodynamic simulations and the ones that are too faint to be seen in current observations. 
This is achieved by coupling the SAM with the \citetalias{Kubota2018} physical AGN spectral model, which outputs physically-motivated SEDs from a model with three distinct radiating zones near SMBHs, dependent on black hole mass and accretion rate, with a number of assumptions regarding the geometry and conditions within these zones. 
By combining the SAM prediction for the distributions of BH masses and AGN accretion rates at different cosmic times with these SED models, we compute the cosmic budget for both hydrogen and helium ionizing photons. This is then combined with an analytic reionization model to forward model the cosmic reionization history.

By making use of three physically motivated models that operate at vastly different physical scales, we  connect the `ground-level', small-scale black hole physics to the `top-level' cosmological observables, such as AGN luminosity functions and the IGM reionization history across cosmic time. 
This gives a comprehensive \textit{overview} of the set of physics derived from observing the nearby Universe and simulations, and provides an \textit{outlook} on the objects forming in the high-redshift Universe to be detected with the next generation of instruments.

We presented predictions for AGN populations at $z = 2$ to 7 with the black hole growth feedback models embedded in the well-established Santa Cruz SAM. We found that the recently published version of these models from \citet{Somerville2021} and the Yung et al. `semi-analytic forecasts for JWST' series produce very good agreement with the hard X-ray luminosity function.

We conducted a controlled experiment and found that increasing the scatter in the $m_\text{BH}$--$m_\text{bulge}$ relation from the value calibrated at $z\sim0$ yields more luminous AGN at $z\gtrsim2$ and improves the agreement with the bright end of the observed AGN luminosity functions between $z \sim 2$ to 4.

We constructed an extended modelling pipeline to link the predicted AGN populations to the subsequent reionization history of intergalactic hydrogen and helium. This is done by forward modelling the AGN SEDs based on the predicted black hole mass and accretion rate in halos across a wide mass range. We provided predictions for the helium ionizing emissivity by summing over the contribution from all predicted AGN. We then tracked the progression of the ionized volume fraction for intergalactic helium utilizing an analytic reionization model. Our models enable self-consistent investigation of the evolution in the AGN populations across cosmic time and their impacts on the cosmic environment.

This work adds new capabilities of predicting spectra and photometry for AGN, as well as their underlying physical properties to a larger \emph{forecasting} framework, which provides tailored predictions for high-redshift observations with \emph{JWST} and other instruments.

We summarize our main conclusions below:

\begin{enumerate}

    \item Using an AGN model that is empirically calibrated and well-tested at low redshifts in conjunction with a physical AGN spectral model, we provide predictions for AGN LFs, helium ionizing photon production rate, and ionized volume fraction between $z = 2$ to 7.

    \item We show that empirical relations that are commonly used to link different observable quantities neglect significant scatter that arises from variation in the physical properties of AGN. 

    \item Our physical model predicts a factor of ten higher helium ionizing emissivity compared to many past studies. We show that this is due to the assumption in previous studies that AGN that are missing from observed UV luminosity functions due to attenuation or obscuration do not contribute to the ionizing photon budget.

    \item Our self-consistent calculations demonstrate that hard ionizing photons produced by accreting SMBHs between $z = 2$ to 7 are sufficient to fully reionize the intergalactic helium by $z\sim2$, with a moderate assumed escape fraction of $\sim 10\%$. This roughly agrees with the fraction of Compton-thin, UV-unattenuated AGN estimated from AGN surveys.

    \item According to our predictions, AGN contribute less than ten percent of the hydrogen ionizing photons at $z>6$.

    \item We provide scaling relations for \NionHeII\ as a function of $M_\text{h}$ throughout the epoch of helium reionization, which may be useful for cosmological-scale reionization simulations where baryonic physics is not resolved. 

    \item We project that AGN of $\log L_\text{bol}$/L$_\odot=10$ to 11 are responsible for producing most of the ionizing photons at $z\sim3$ to 4, during the epoch where intergalactic helium is getting reionized most rapidly.

    \item The contribution of ionizing photons from AGN increases rapidly at $z<6$, until they are contributing an equal or slightly larger amount than star forming galaxies by $z\sim 2$. This may help to sustain hydrogen ionization at late times, in a scenario where galaxy escape fractions turn out to be extremely low (less than a few percent).  

\end{enumerate}

\section*{Acknowledgements}
The authors of this paper would like to thank M\'elanie Habouzit and Xuejian Shen for providing access to their work and for the many insightful discussions.
We also thank Joel Primack, Sangeeta Malhotra, James Rhoads, Sultan Hassan, and Greg Bryan for useful discussions. 
AY thanks the organizers of the 2020 `Summer All Zoom Epoch of Reionization Astronomy Conference (SAZERAC)', which provided a platform for talks and discussions that inspired this work. 
AY is grateful for the hospitality of the Flatiron Institute during the time of creating this work.
We also thank the referee Xuejian Shen for constructive comments that improved this paper.
The simulations and data products for this work were run and stored on computing cluster \emph{rusty} at the Center for Computational Astrophysics, Flatiron Institute. 
AY and RSS thank the Downsbrough family for their generous support, and gratefully acknowledge funding from the Simons Foundation. 
AY is supported by an appointment to the NASA Postdoctoral Program (NPP) at NASA Goddard Space Flight Center, administered by Universities Space Research Association under contract with NASA.

\section*{Data Availability}
The data underlying this paper are available in the Data Product Portal hosted by the Flatiron Institute at  \url{https://www.simonsfoundation.org/semi-analytic-forecasts-for-jwst/}.



\bibliographystyle{mnras}
\bibliography{library.bib}



\appendix

\section{AGN UV attenuation and obscuration}
\label{appendix:a}
\setcounter{table}{0} \renewcommand{\thetable}{A\arabic{table}}

\begin{figure}
    \includegraphics[width=\columnwidth]{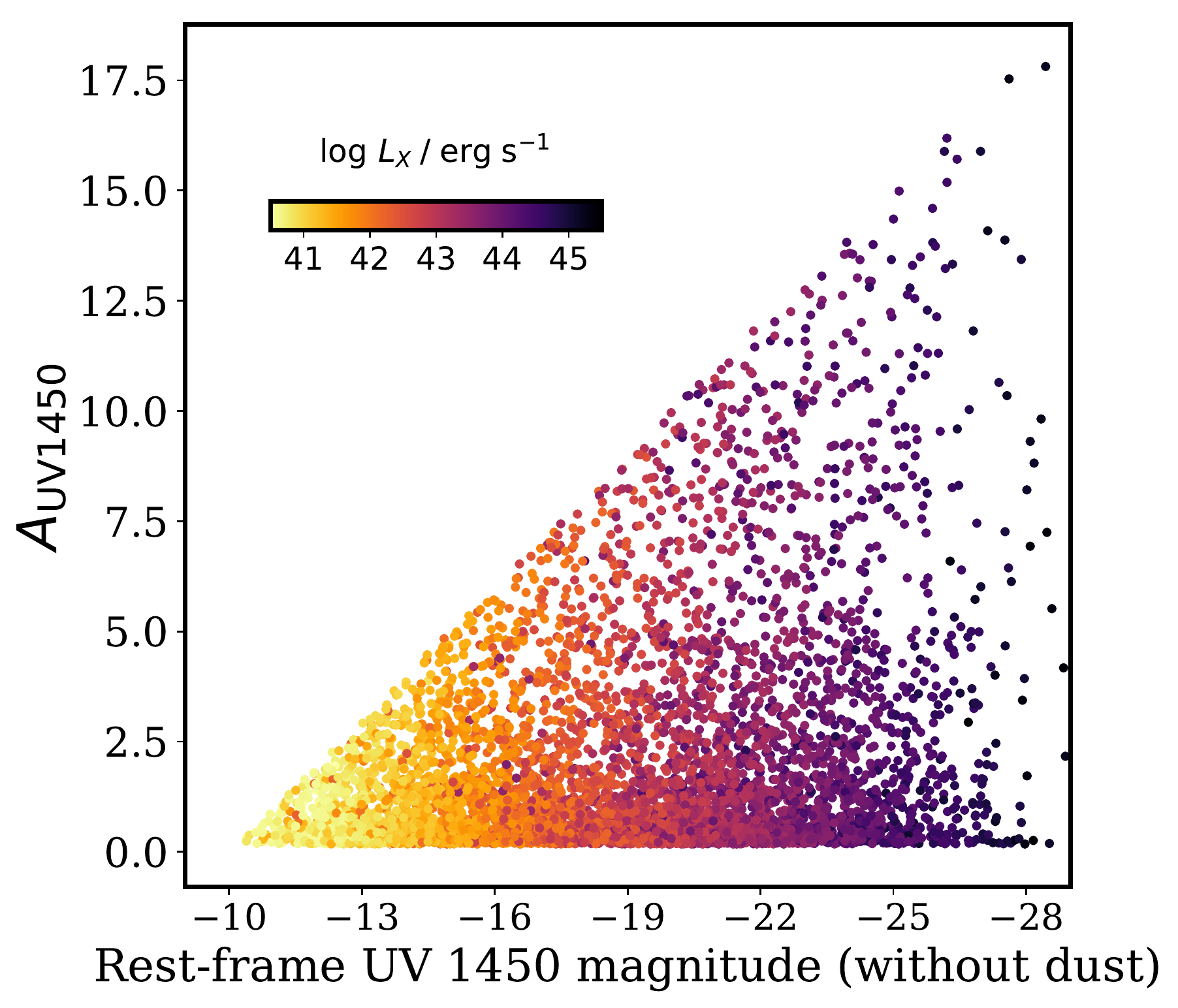}
    \caption{Distribution of intrinsic rest-frame UV1450 magnitude versus UV-band attenuation for our simulated AGN population. Here we show all AGN predicted across $z = 2$ -- 7 since none of the model components used in this calculation explicitly depend on redshift. 
    The data points are colour-coded to indicate their intrinsic hard X-ray luminosity.}
    \label{fig:UV1450_AUV} 
\end{figure}

In this Appendix, we provide an additional diagnostic diagram, Fig.~\ref{fig:UV1450_AUV}, for UV attenuation applied to our AGN based on the prescriptions described in Section \ref{sec:UVLF}. This relation is not expected to evolve with time as none of the model components involved in this calculation explicitly depends on redshift. We also note that the upper limit of this relation is inherited from the fixed ratio that converts neutral hydrogen to colour excess. 
The data points in the scatter plot are colour-coded for the X-ray luminosity $L_x$ of predicted AGN, which is one of the variables that influences the prediction of $N_\text{H}$ in the radiation-lifted torus model proposed by \citet{Buchner2017}.
This figure is comparable to a similar figures for galaxies presented in fig.~4 in \citetalias{Yung2019}.

\section{Extended Predictions on UV luminosity functions}
\label{appendix:b}
\setcounter{table}{0} \renewcommand{\thetable}{B\arabic{table}}

\begin{figure}
    \includegraphics[width=\columnwidth]{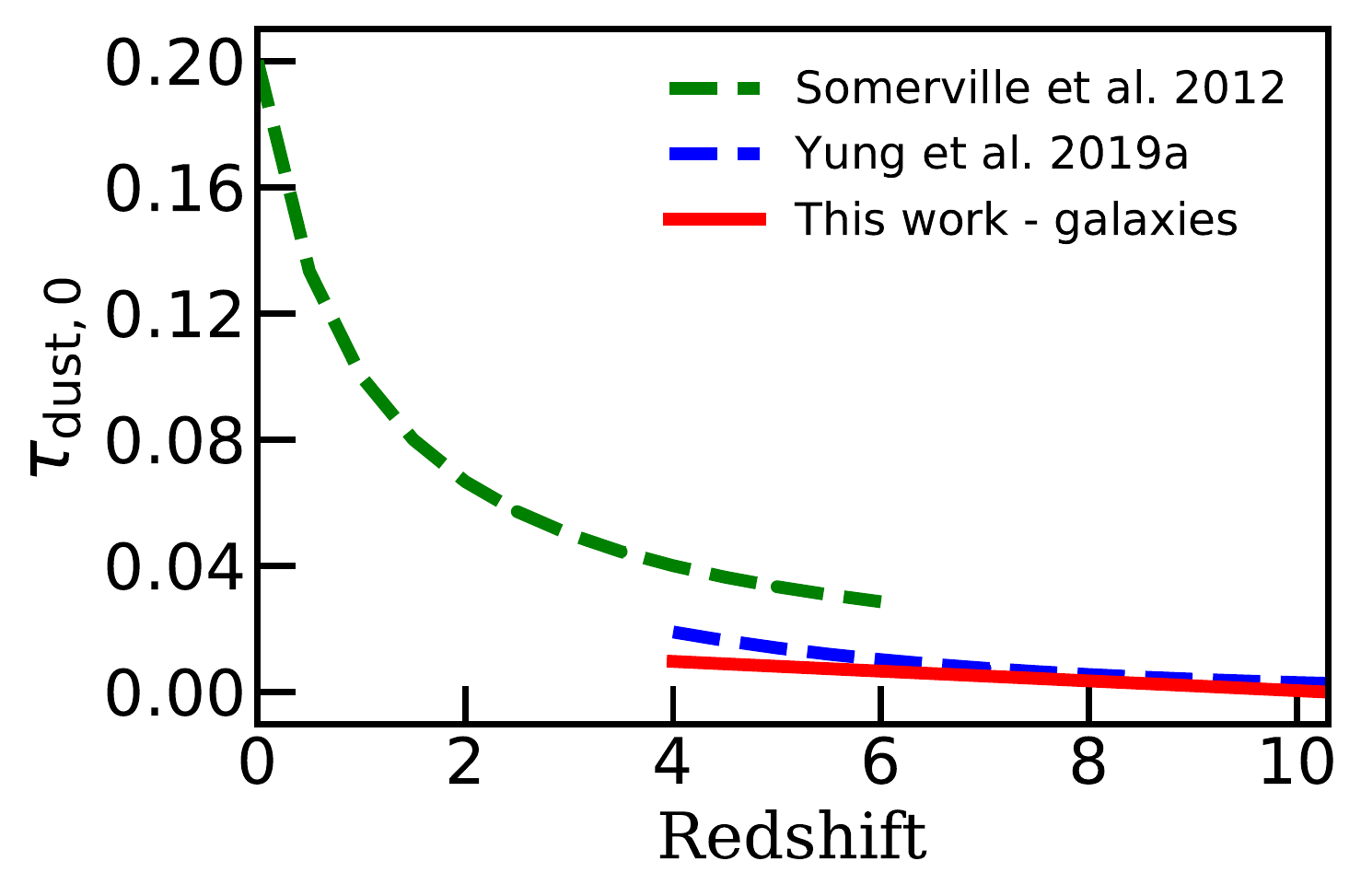}
    \caption{The $\tau_\text{dust,0}(z)$ values required to reproduce observations in the current model (red solid line), compared to the values adopted in \citet[][green dashed line]{Somerville2012} and \citet[][blue dashed line]{Yung2019}.}
    \label{fig:tau_dust_0}
\end{figure}

Upon updating the black hole growth model to include the feeding of cold gas via disc instability, we are also predicting some stronger AGN feedback onto the galaxies forming in larger halos. As a result of that, fewer bright, massive galaxies have formed, and therefore the amount of attenuation by dust required to reproduce the observed UV LFs have gone down slightly. The re-calibrated redshift evolution of the redshift-dependent dust optical depth is $\tau_\text{dust,0} = -0.001543z + 0.01583$. See section 2.6 in \citetalias{Yung2019} for full description for our dust attenuation model. We compare this updated parameter with previous work in Fig.~\ref{fig:tau_dust_0} \citep{Somerville2012, Yung2019}.

\begin{figure}
    \includegraphics[width=\columnwidth]{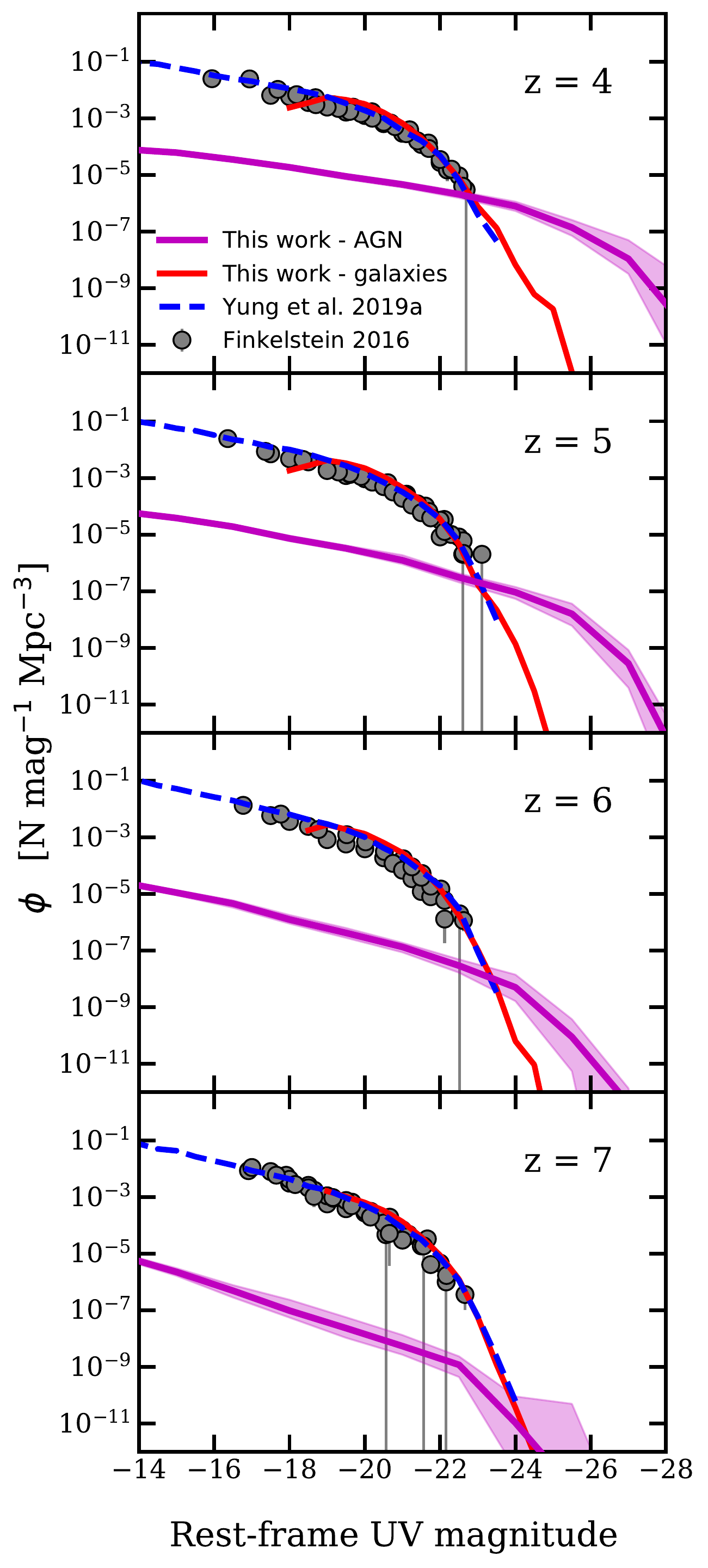}
    \caption{Rest-frame UV LFs from our fiducial model for galaxies predicted in this work (red, $V_\text{vir} = 100$ -- 1400\kms) and from previous works in the paper series (blue, $V_\text{vir} = 20$ -- 500\kms). We also include a compilation of observational constraints from \citet[][circles]{Finkelstein2016} as shown in \citetalias{Yung2019}. We also include the AGN UVLFs predicted in this work for comparison (magenta line and shaded area, see Fig.~\ref{fig:Compare_UV1450_dust_LFs} and associated text for detail).}
    \label{fig:UV_LFs_extended}
\end{figure}

We present rest-frame UV (1600\AA) luminosity functions at $z = 4$ to 7 from the set of predictions shown in this work (red, forming in root halos spanning $V_\text{vir} = 100$ -- 1400\kms), compared to predictions from previous work (blue, $V_\text{vir} = 20$ -- 500\kms). These luminosities only account for radiation from stellar populations but not AGNs. 
Both sets of predictions share an identical `fiducial' model configurations (see \citetalias{Yung2019} for detail). The finer grid of root halos behind these new predictions provides more reliable sampling for the population of bright galaxies forming in massive halos. Note that massive objects at this redshift range are extremely rare and they are unlikely to be picked up by the relatively small field of view of \textit{JWST}. These results are shown in Fig.~\ref{fig:UV_LFs_extended} and tabulated values in Table \ref{table:extended_UVLF_dust}.

\begin{table}
    \centering
    \caption{Tabulated UV LFs at $z = 4$--7 predicted for new halo mass range as presented in Fig.~\ref{fig:UV_LFs_extended}.}
    \label{table:extended_UVLF_dust}
    \begin{tabular}{ccccc} 
        \hline
        $M_\text{UV}$ & $z=4$ & $z=5$ & $z=6$ & $z=7$      \\
        \hline
        $-25.5$ & $-11.97$ & ---      & ---      & ---     \\
        $-25.0$ & $-9.73$  & $-12.78$ & ---      & ---     \\
        $-24.5$ & $-9.22$  & $-10.53$ & $-11.03$ & $-12.08$\\
        $-24.0$ & $-8.18$  & $-8.86$  & $-10.20$ & $-10.44$\\
        $-23.5$ & $-6.87$  & $-7.66$  & $-8.33$  & $-8.92$ \\
        $-23.0$ & $-6.12$  & $-6.78$  & $-6.98$  & $-7.25$ \\
        $-22.5$ & $-5.12$  & $-5.35$  & $-5.75$  & $-5.91$ \\
        $-22.0$ & $-4.36$  & $-4.46$  & $-4.81$  & $-5.06$ \\
        \hline
    \end{tabular}
\end{table}

\section{Comparison to empirical conversion}
\label{appendix:c}
\setcounter{table}{0} \renewcommand{\thetable}{C\arabic{table}}

\begin{figure*}
    \includegraphics[width=1.7\columnwidth]{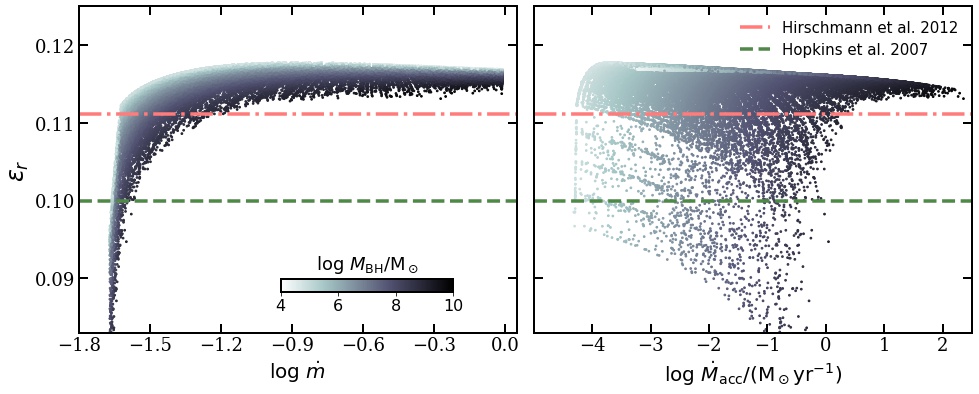}
    \caption{\textit{Left panel:} AGN radiative efficiency $\epsilon_r$ versus Eddington-normalized accretion rate, \mdot, colour-coded for \mBH, assuming isotropic emission (inclination $45^{\circ}$). This distribution is compared to the constant values adopted in \citet{Hopkins2007} and \citet{Hirschmann2012a}. 
    \textit{Right panel:} $\epsilon_r$ versus accretion rate in physical units, colour-coded for \mBH. These figures illustrate the large scatter in radiative efficiency that is not captured by commonly used empirical conversions. }
    \label{fig:Lbol_diagnostics}
\end{figure*}

In Fig.~\ref{fig:Lbol_diagnostics}, we compare the radiative efficiency $\epsilon_r \equiv L_\text{bol}/(\dot{M}_\text{acc} c^2)$ from our predictions with the typical values adopted in past studies \citep{Hopkins2007, Hirschmann2012a}. 
While some $z\sim0$ observations indicate that an acceptable range would be $0.04 < \epsilon_r < 0.16$ \citep{Marconi2004},
this work demonstrates how this quantity varies as a function of BH mass and accretion rate.

\section{Selected Tabulated Data}
\label{appendix:d}
\setcounter{table}{0} \renewcommand{\thetable}{D\arabic{table}}
We provide tabulated data for AGN LFs and ionizing emissivity as a function of halo mass in Table~\ref{table:AGN_LF} and Table~\ref{table:NHe_mhalo}. The full range of tabulated data is available online in our data release portal.

\begin{table}
    \centering
    \caption{Tabulated AGN LFs at $z = 2$--7 predicted for new halo mass range.}
    \label{table:AGN_LF}
    \begin{tabular}{ccccccc} 
        \hline
        $\log(L_\text{bol}/\text{L}_\odot)$ & $z=2$ & $z=3$ & $z=4$ & $z=5$ & $z=6$ & $z=7$\\ 
        \hline
        $9.0$ & $-3.61$ & $-3.05$ & $-2.81$ & $-2.78$ & $-3.03$ & $-3.29$\\
        $9.5$ & $-3.32$ & $-2.91$ & $-2.78$ & $-2.89$ & $-3.21$ & $-3.65$\\
        $10.0$ & $-3.25$ & $-3.01$ & $-2.97$ & $-3.11$ & $-3.54$ & $-4.19$\\
        $10.5$ & $-3.45$ & $-3.22$ & $-3.23$ & $-3.52$ & $-4.05$ & $-4.91$\\
        $11.0$ & $-3.65$ & $-3.47$ & $-3.55$ & $-3.89$ & $-4.58$ & $-5.53$\\
        $11.5$ & $-3.85$ & $-3.75$ & $-3.87$ & $-4.26$ & $-5.30$ & $-6.31$\\
        $12.0$ & $-4.18$ & $-4.04$ & $-4.23$ & $-4.81$ & $-5.65$ & $-7.22$\\
        $12.5$ & $-4.47$ & $-4.42$ & $-4.59$ & $-5.41$ & $-6.35$ & $-7.76$\\
        $13.0$ & $-4.74$ & $-4.82$ & $-5.03$ & $-5.87$ & $-7.61$ & $-9.33$\\
        $13.5$ & $-5.21$ & $-5.26$ & $-5.80$ & $-6.66$ & $-8.61$ & ---\\
        $14.0$ & $-5.72$ & $-6.36$ & $-6.96$ & $-8.38$ & --- & ---\\
        $14.5$ & $-6.38$ & $-7.58$ & $-8.03$ & --- & --- & ---\\
        $15.0$ & $-8.89$ & $-8.39$ & --- & --- & --- & ---\\
        \hline
    \end{tabular}
\end{table}

\begin{table*}
    \centering
    \caption{Tabulated $\log(\dot{N}_\text{ion,He}/M_\text{h})$ at $z = 2$--7 predicted for new halo mass range.}
    \label{table:NHe_mhalo}
    \begin{tabular}{ccccccc} 
        \hline
         & \multicolumn{6}{c}{$\log((\dot{N}_\text{ion,He}/M_\text{h})/(\text{s M}_\odot)^{-1})$} \\
        $\log(M_\text{h}/\text{M}_\odot)$ & $z=2$ & $z=3$ & $z=4$ & $z=5$ & $z=6$ & $z=7$ \\
        \hline
        11.00 & 52.47$^{+0.75}_{-0.80}$ & 52.20$^{+0.85}_{-0.86}$ & 52.01$^{+0.83}_{-0.81}$ & 52.06$^{+0.76}_{-0.85}$ & 52.11$^{+0.70}_{-0.78}$ & 52.08$^{+0.60}_{-0.58}$ \\ [0.05in]
        11.50 & 52.82$^{+0.99}_{-0.95}$ & 52.78$^{+1.07}_{-1.04}$ & 52.61$^{+1.08}_{-1.09}$ & 52.57$^{+0.98}_{-1.11}$ & 52.60$^{+0.83}_{-0.93}$ & 52.58$^{+0.73}_{-0.63}$ \\ [0.05in]
        12.00 & 53.97$^{+0.94}_{-1.24}$ & 53.94$^{+1.00}_{-1.45}$ & 53.52$^{+1.18}_{-1.27}$ & 53.39$^{+1.03}_{-1.10}$ & 53.38$^{+0.82}_{-0.84}$ & 53.22$^{+0.89}_{-0.76}$ \\ [0.05in]
        12.50 & 55.04$^{+0.66}_{-0.99}$ & 54.70$^{+0.80}_{-1.25}$ & 54.36$^{+0.93}_{-1.79}$ & 53.90$^{+1.04}_{-1.18}$ & 53.90$^{+0.90}_{-0.90}$ & 53.80$^{+0.75}_{-0.88}$ \\ [0.05in]
        13.00 & 55.64$^{+0.38}_{-1.15}$ & 55.30$^{+0.60}_{-0.97}$ & 54.75$^{+0.77}_{-1.57}$ & 54.27$^{+0.98}_{-1.56}$ & 54.03$^{+1.00}_{-1.12}$ & 53.84$^{+1.00}_{-0.93}$ \\ [0.05in]
        13.50 & 55.74$^{+0.55}_{-1.06}$ & 55.60$^{+0.60}_{-0.93}$ & 55.30$^{+0.58}_{-1.23}$ & 54.46$^{+0.94}_{-1.59}$ & 54.08$^{+1.25}_{-1.38}$ & 54.00$^{+0.85}_{-1.03}$ \\ [0.05in]
        14.00 & 56.10$^{+0.38}_{-0.85}$ & 55.65$^{+0.65}_{-0.76}$ & 55.35$^{+0.73}_{-1.21}$ & 54.85$^{+0.74}_{-1.44}$ & 53.93$^{+1.17}_{-1.35}$ & 53.17$^{+1.60}_{-0.67}$ \\ [0.05in]
        \hline
    \end{tabular}
\end{table*}

\bsp	
\label{lastpage}
\end{document}